\documentclass{CUP-JNL-DCE}

\usepackage{ifpdf}
\usepackage[T1]{fontenc}
\usepackage{sourcesanspro}
\usepackage{textcomp}
\usepackage{xcolor}
\usepackage{hyperref}

\usepackage[utf8]{inputenc}
\usepackage{textcomp}
\usepackage{graphicx}
\usepackage{subfigure}
\usepackage{color}
\usepackage{tabularx}
\usepackage{longtable}
\usepackage{footmisc} 
\usepackage[acronyms]{glossaries}
\makenoidxglossaries
\usepackage{listings}
\definecolor{numb}{RGB}{25,75,67}
\definecolor{string}{rgb}{0.64,0.08,0.08}
\lstdefinestyle{python}{
    numbers=none,
    frame=single,
    rulecolor=\color{black},
	showstringspaces=false
    showspaces=false,
    showtabs=false,
    breaklines=true,
    postbreak=\raisebox{0ex}[0ex][0ex]{\ensuremath{\color{gray}\hookrightarrow\space}},
    breakatwhitespace=true,
    basicstyle=\ttfamily\small,
    basewidth=4.0pt,
    linewidth=410pt, 
    upquote=true,
}
\lstdefinelanguage{docker}{
  keywords={FROM, RUN, COPY, ADD, ENTRYPOINT, CMD,  ENV, ARG, WORKDIR, EXPOSE, LABEL, USER, VOLUME, STOPSIGNAL, ONBUILD, MAINTAINER, HEALTHCHECK},
  keywordstyle=\color{black}\bfseries,
  identifierstyle=\color{black},
  sensitive=false,
  comment=[l]{\#},
  commentstyle=\color{black}\ttfamily,
  stringstyle=\color{black}\ttfamily,
  morestring=[b]',
  morestring=[b]"
}
\newcommand{\figref}[1]{\figurename~\ref{#1}}

\newcommand{\req}[2]{\newline\noindent\hypertarget{#1}{\textbf{#1: #2}}}
\newcommand{\reqref}[1]{\hyperlink{#1}{#1}}
\newcommand{\ttinnewline}[1]{\noindent\texttt{#1}\\\noindent}

\newcommand{\lstkw}[1]{\lstinline|#1|}

\newcommand{\blinded}[2]{#1}

\articletype{RESEARCH ARTICLE}
\jname{Data-Centric Engineering}
\jyear{2024}

\newacronym{AC}{AC}{Air Conditioning}
\newacronym{API}{API}{Application Programming Interface}
\newacronym{BAS}{BAS}{Building Automation System}
\newacronym{BSS}{BSS}{Battery Storage System}
\newacronym{CPU}{CPU}{Central Processing Unit}
\newacronym{EM}{EM}{Energy Management}
\newacronym{BEOA}{BEOA}{Building Energy Optimization Algorithm}
\newacronym{EMS}{EMS}{Energy Management System}
\newacronym{ESG}{ESG}{Energy Service Generics}
\newacronym{GUI}{GUI}{Graphical User Interface}
\newacronym{HAL}{HAL}{Hardware Abstraction Layer}
\newacronym{HTTP}{HTTP}{Hypertext Transfer Protocol}
\newacronym{HTTPS}{HTTPS}{Hypertext Transfer Protocol Secure}
\newacronym{HVAC}{HVAC}{Heating, Ventilation and Air Conditioning}
\newacronym{IdP}{IdP}{Identity Provider}
\newacronym{IPC}{IPC}{Inter-Process Communication}
\newacronym{JSON}{JSON}{JavaScript Object Notation}
\newacronym{JWT}{JWT}{JSON Web Token}
\newacronym{MBRL}{MBRL}{Model Based Reinforcement Learning}
\newacronym{ML}{ML}{Machine Learning}
\newacronym{MPC}{MPC}{Model Predictive Control}
\newacronym{NN}{NN}{Neural Network}
\newacronym{OIDC}{OIDC}{OpenID Connect}
\newacronym{PV}{PV}{Photovoltaic}
\newacronym{RPC}{RPC}{Remote Procedure Call}
\newacronym{REST}{REST}{Representational State Transfer}
\newacronym{RL}{RL}{Reinforcement Learning}
\newacronym{RBC}{RBC}{Rule Based Control}
\newacronym{URL}{URL}{Uniform Resource Locator}

\begin{document}

\begin{Frontmatter}

\title[Open Energy Services]
{Open Energy Services - Forecasting and Optimization as a Service for Energy Management Applications at Scale}

\author*[1]{David Wölfle}\orcid{0009-0004-6249-930X}\email{woelfle@fzi.de}
\author[2]{Kevin Förderer}\orcid{0000-0002-9115-670X}
\author[1]{Tobias Riedel}\orcid{0009-0001-2335-7278}
\author[2]{Natascha Fernengel}\orcid{0009-0009-9473-186X}
\author[1]{Lukas Landwich}
\author[2]{Ralf Mikut}\orcid{0000-0001-9100-5496}
\author[2]{Veit Hagenmeyer}\orcid{0000-0002-3572-9083}
\author[1]{Hartmut Schmeck}\orcid{0000-0002-4295-7631}

\authormark{David Wölfle \textit{et al.}}

\address[1]{\orgdiv{Intelligent Systems and Production Engineering}, \orgname{FZI Research Center for Information Technology}, \orgaddress{\street{Haid-und-Neu-Str. 10–14},\postcode{76131 Karlsruhe}, \country{Germany}}}

\address*[2]{\orgdiv{Institute for Automation and Applied Informatics}, \orgname{Karlsruhe Institute of Technology}, \orgaddress{\street{Hermann-von-Helmholtz-Platz 1}, \postcode{76344 Eggenstein-Leopoldshafen}, \country{Germany}}}

\received{05 August 2024}


\abstract{
This article aims at facilitating the widespread application of \glspl{EMS}, especially on buildings and cities, in order to support the realization of future carbon-neutral energy systems.
We claim that economic viability is a severe issue for the utilization of EMSs at scale and that the provisioning of forecasting and optimization algorithms as a service can make a major contribution to achieve it.
To this end, we present the \emph{Energy Service Generics} software framework that allows the derivation of fully functional services from existing forecasting or optimization code with ease.
This work documents the strictly systematic development of the framework, beginning with a requirement analysis, from which a sophisticated design concept is derived, followed by a description of the implementation of the framework.
Furthermore, we present the concept of the \emph{Open Energy Service} community, our effort to continuously
maintain the service framework but also provide ready-to-use forecasting and optimization services.
Finally, an evaluation of our framework and community concept, as well as a demarcation between our work and the current state of the art, is presented.
}

\begin{policy}[Impact Statement]
Energy management will likely play a vital role in future carbon-neutral energy systems, as it allows for unlocking energy efficiency and flexibility potentials.
However, energy management systems need to be applied at large scales to realize the desired effect, which clearly requires minimization of costs for setup and operation.
We promote an approach to split the complex optimization algorithms employed by energy management systems into standardized components, which can be provided as a service with marginal costs at scale.
This work introduces a framework as well as a community concept to support the efficient implementation and operation of such services.
Thus, this work is a significant step towards the large-scale application of energy management systems aiding a carbon-neutral future.
\end{policy}

\end{Frontmatter}

\section{Introduction} \label{sec_introduction}
Global scale efforts are required to mitigate the most severe consequences of climate change, including a significant increase in the energy efficiency of consumers as well as the decarbonization of energy supply \cite{shukla_summary_2022}.
The vast utilization of renewable energy sources required for the latter will additionally likely induce an increased demand for energy flexibility by consumers \cite{alizadeh_flexibility_2016,kondziella_flexibility_2016,papaefthymiou_towards_2016}.
\glspl{EMS}, in a sense of software computing optimized operational schedules and executing these on devices and systems, have been demonstrated to be capable of reducing energy demand, lowering $CO_2$ emissions and/or unlocking flexibility \cite{schibuola_demand_2015,oldewurtel_use_2012,ding_octopus_2019,salpakari_optimal_2016,chen_gnu-rl_2019}.
However, in order to achieve the desperately needed global impact energy management solutions will be required at scale, like e.g. applied to thousands of buildings.

Economic viability is certainly a key factor for the widespread adoption of \glspl{EMS}.
Forecasting and optimization algorithms are essential parts of \glspl{EMS} (see Section \ref{sec_nomenclature}), but have traditionally been developed for a single specific target (e.g. for one particular building) \cite{wolfle_guide_2020}, like in  \cite{schibuola_demand_2015,oldewurtel_use_2012,ding_octopus_2019,salpakari_optimal_2016,chen_gnu-rl_2019} or the publications reviewed by \cite{shaikh_review_2014}.
This approach is problematic as it has been shown that the development costs of target-specific forecasting and optimization algorithms are higher than the monetary savings, even for medium-sized commercial buildings \cite{gwerder_final_2013}.

This paper aims at supporting the widespread adoption of \glspl{EMS} by enabling the utilization of forecasting and optimization algorithms for energy management applications at large scales.
Our approach, as discussed in Section \ref{sec_nomenclature} in detail, is to replace target-specific forecasting and optimization algorithms (which are locally deployed as part of the \gls{EMS} instances) with generic forecasting and optimization algorithms that are centrally provided as web services, in order to reduce development and operation costs of \glspl{EMS}.
In Section \ref{sec_related_work} we extensively analyze the current state of the art and find that the concept of providing forecasting and optimization algorithms as web services is already well established, especially in commercial solutions provided by international corporations.
Furthermore, it is relevant to note that data-driven algorithms, i.e. forecasting and optimization approaches generally suitable for utilization in larger scales of \glspl{EMS} controlling heterogeneous systems, have been frequently proposed in academia \cite{anand_bottom-up_2023,chen_gnu-rl_2019,ding_octopus_2019,meisenbacher_autopv_2023,xuereb_conti_physics-based_2023}.
However, it seems that currently no software framework exists that supports the implementation and operation of such services,
which seems to be a major barrier for bringing these new and innovative forecasting and optimization algorithms into practical application by \glspl{EMS}.

The present paper addresses the aforementioned shortcoming by contributing a \emph{framework} that allows the provisioning of forecasting or optimization code as a web service.
To that end, we begin by carrying out an extensive analysis to specify requirements (Section~\ref{sec_requirements}).
Based on this, we present a sophisticated design concept that satisfies these requirements (Section~\ref{sec_design_concept}) and finally derive an implementation of our concept (Section~\ref{sec_implementation}), which we release as a free and open source repository alongside this publication.
Our second contribution is the presentation of our concept for the \emph{Open Energy Services} community (Section~\ref{sec_community_concept}), a group that is dedicated to the maintenance of the framework, but also to the development and operation of forecasting and optimization services.
Finally, Section \ref{sec_evaluation} is devoted to demonstrating that our contributions, i.e. framework and community concept, are useful for facilitating the development and operation of forecasting and optimization services for energy management applications.
\section{Nomenclature}\label{sec_nomenclature}
\begin{figure}
\centering
\FIG{\includegraphics[width=1.0\columnwidth]{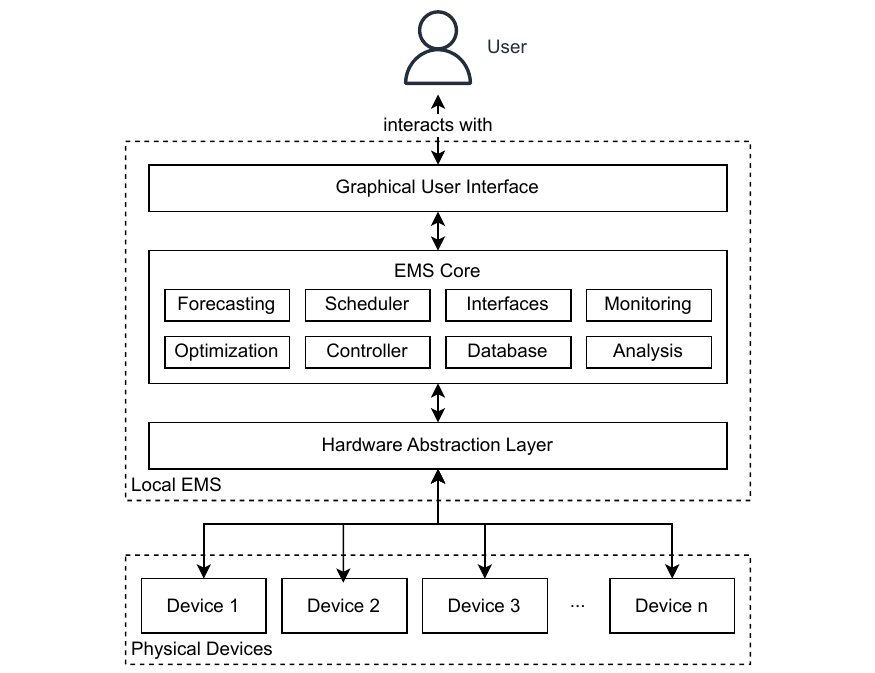}}{\caption{Typical high-level architecture of an \gls{EMS}}\label{fig_typical_ems_architecture}}
\vspace*{-0.4cm}
\end{figure}

As a first step to define the context this work is set in, we begin with inspecting the typical \acrfull{EMS} architecture\footnote{It is worth noting that many commercial offerings of \glspl{EMS} are regularly reduced to an implementation of ISO 50001, which defines energy management applications with a focus on monitoring and analysis. Such \glspl{EMS} will usually not contain forecasting and optimization components. Thus, they are not referenced here as they are not covered by the scope of this work. However, the service approach promoted in this work might in fact be a relevant option for unlocking to potential of forecasting and optimization algorithms for such \glspl{EMS}.}.
Concrete proposals for the latter have been provided by \cite{dawson-haggerty_boss_2013,lee_design_2016,mauser_organic_2015,pipattanasomporn_bemoss_2015}. \cite{han_home_2023} contains a review of architectures of \gls{EMS} for residential buildings.
While these papers generally show no consensus about the internal structures of \gls{EMS}, it is nevertheless easily possible to map the respective suggested architectures to the convention introduced below and summarized in \figref{fig_typical_ems_architecture}. 
The latter also holds for the architecture of OpenEMS\footnote{\href{https://openems.github.io/openems.io/openems/latest/edge/architecture.html}{https://openems.github.io/openems.io/openems/latest/edge/architecture.html}}, the only \gls{EMS} the authors are aware of that is developed as an open source project by a consortium of commercial institutions.

In order to discuss the internal structures of \glspl{EMS} in greater detail, we consider a running example of a commercial building equipped with a \gls{PV} and \gls{BSS} as \emph{physical devices} as well as an \gls{EMS}, intended to optimize the operation of the latter, that is executed on a computing device inside the building.
As the facility managers of the building are responsible for its correct operation, they interact with the \gls{EMS}, e.g. to monitor the operation or to adjust setpoints.
However, other residents of the building may interact with the \gls{EMS} too, e.g. to specify their personal demands which the system should consider.
Thus, all persons interacting with the \gls{EMS} are the \emph{users} of it.
The \gls{EMS} has been developed and is supported by a specialized institution, the \emph{\gls{EMS} developer}.

The \gls{EMS} itself is essentially a piece of software consisting of three major parts:
\begin{enumerate}
	\item A \emph{\gls{GUI}} which the users interact with.
	\item A \emph{\gls{HAL}} connecting the \gls{EMS} to the physical devices.
	\item A component holding the essential management functionality, which we will refer to in this work as \emph{\gls{EMS} core}.
\end{enumerate}
Returning to our running example, we can perceive the functionality of the \gls{EMS} core part to contain:
\begin{itemize}
	\item \emph{Optimization}: Computes optimized \emph{schedules} for the controllable devices in order to satisfy the goals provided by the users. E.g. the facility manager could configure the \gls{EMS} such that the \gls{BSS} is used to take advantage of flexible electricity tariffs. 
	\item \emph{Forecasting}: Computes \emph{forecasts} that the optimization algorithm requires as input. In the present example, the optimization could require predictions of the future development of the energy price, the electric load, and the power generation of the \gls{PV} system.
	\item \emph{Scheduler}: Invokes the forecasting and optimization algorithms periodically or at certain events. In the present example, the scheduler might trigger the computation of an optimized schedule for the \gls{BSS} every 15 minutes by first invoking the forecasting algorithms and then forwarding the predictions (along with any other required input data) to the optimization algorithm. The scheduler might additionally fetch data from external sources, like e.g. a weather forecast as necessary input for a \gls{PV} power prediction algorithm.
	\item \emph{Controller}: Ensures that the user and hardware constraints are satisfied by the \gls{EMS}. For example, the facility manager could wish to enforce that the \gls{BSS} is not discharged below 20\% in order to expand the lifetime of the device. The controller might additionally contain simple rules that define a sane default strategy in case that the optimization algorithm does not work as intended.
	\item \emph{Database}: Stores the data required for the operation of the \gls{EMS} (incl. \gls{GUI}), like, for example, measurements emitted by the physical devices.
	\item \emph{Interfaces}: Implements the connectivity to the \gls{GUI} and the \gls{HAL}. Might additionally contain interfaces for external applications or a message broker for internal communication between parts of \gls{EMS} core.
	\item \emph{Monitoring}: Continuously oversees the system influenced by the \gls{EMS} and emits alerts in case of malfunctioning. The monitoring system could, for example, send an email to the facility manager if the communication with devices has been lost or these need maintenance.
	\item \emph{Analysis}: Aggregates and computes metrics relevant for the users of the \gls{EMS}, for example statistics about the energy usage pattern.
\end{itemize}

\begin{figure}
\centering
\FIG{\includegraphics[width=1.0\columnwidth]{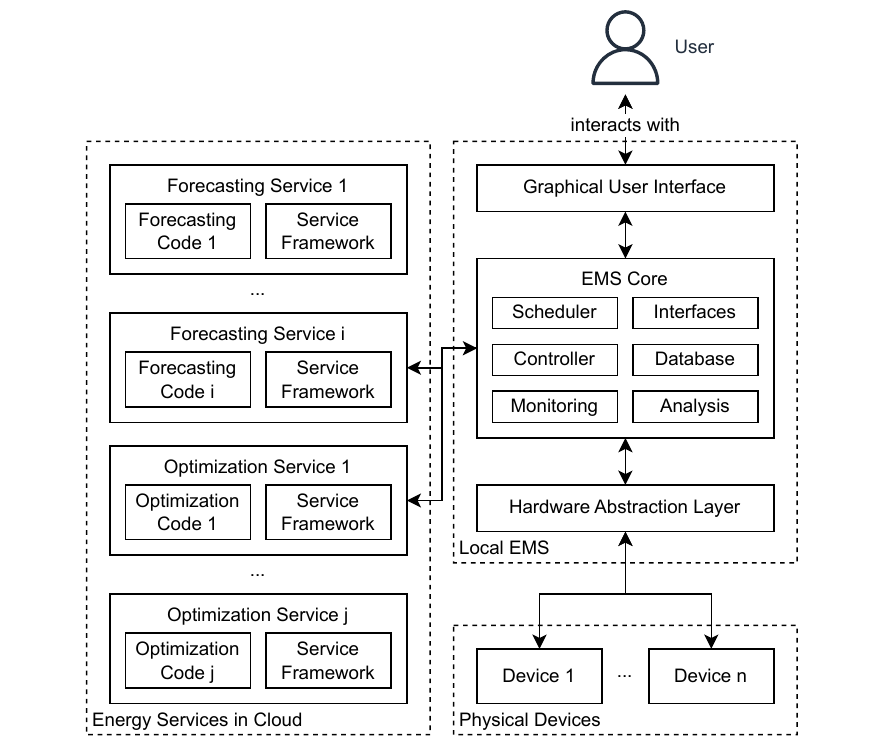}}{\caption{High-level architecture of an \gls{EMS} utilizing selected forecasting and optimization services}\label{fig_ems_architecture_with_services}}
\vspace*{-0.5cm}
\end{figure}

In contrast to the usual \gls{EMS} architecture pattern introduced above, this work promotes an approach in which the forecasting and optimization algorithms are not directly integrated into the \gls{EMS}, but provided as services, as summarized in \figref{fig_ems_architecture_with_services}.
It is worth noting that, although forecasting and/or optimization algorithms are utilized as a service, the correct operation of the \gls{EMS} remains the responsibility of the \gls{EMS} developer, i.e. by implementing a controller component (see above) into the \gls{EMS}.

In the context of this work, a \emph{service} refers to a web-based program that provides a functionality required for energy management applications via a standardized interface.
The intention of providing forecasting and optimization algorithms as services is to make these algorithms available to a larger number of \glspl{EMS} in order to reduce development and maintenance costs of the individual systems.
The separation of forecasting and optimization algorithms from the \gls{EMS} software implies the need to extend the former with interfaces in order to allow the interaction between the services and the \glspl{EMS}.
Furthermore, the intended usage of the forecasting and optimization algorithms by a large number of \glspl{EMS} makes it necessary to consider how these can be executed in a scalable way.
For example, one should consider that the implementation of a forecasting or optimization algorithm, henceforth referred to as the \emph{forecasting or optimization code}, will generally not contain an \gls{API} suitable for web-based clients or functionality to concurrently handle thousands of requests.
Thus, it is necessary to extend the forecasting or optimization code, with all the functionality required for an operation as a service.
However, it is obviously not very effective to develop and implement this extension for every service from scratch, as it will likely be very similar for all services.
Hence, the utilization of a \emph{service framework}, i.e. a software that drastically reduces the necessary effort for developing forecasting and optimization services, by providing the software parts that are generic for all services.
In fact, a large fraction of this work is devoted to the design and implementation of such a service framework.

It should be noted that this service framework is by no means limited to services for forecasting and optimization:
Consider e.g. a heuristic that detects the occupancy in a building from limited information, or an algorithm (like in \cite{de_jongh_physics-informed_2022}) that determines the current state of the electricity grid.
The provisioning of such algorithms as services is clearly useful in the wider sense of energy management.
Furthermore, by its generic design, the proposed framework is applicable for use cases not related to energy management. Consider, for example, the flood prediction approach proposed by \cite{hofmeister_semantic_2024}, which could be provided as a service, too.
However, in the following, for simplicity and readability, we refer to \emph{forecasting and optimization} or to the retrieval of a \emph{forecast or optimized schedule}.
This is not meant to exclude other, not strictly covered, but related algorithms.

Finally, it is necessary to regard the development process of a forecasting and optimization service as well as the corresponding stakeholders that are involved.
\begin{figure}
\centering
\FIG{\includegraphics[width=1.0\columnwidth]{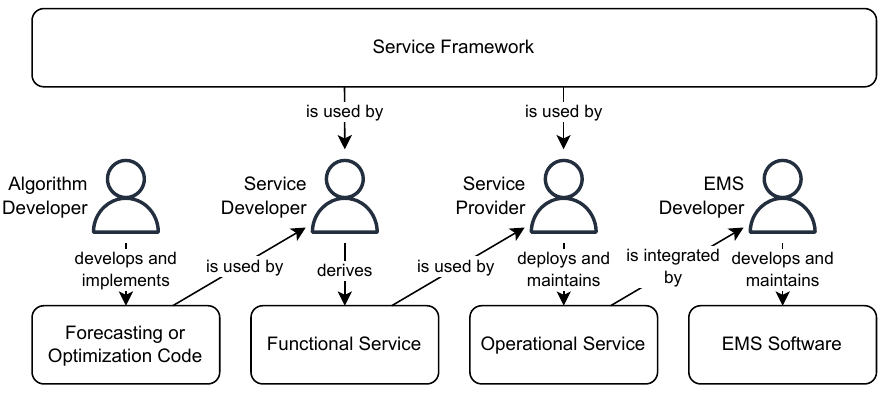}}{\caption{Stakeholders involved in the development process of forecasting or optimization services}\label{fig_service_development_process_and_stakeholders}}
\vspace*{-0.7cm}
\end{figure}
The first step is the development and implementation of the forecasting or optimization algorithm by the \emph{algorithm developer}.
This step might have been finished far before the development of a service has been decided and we thus use the terminology of an \emph{existing forecasting or optimization code}, in order to illustrate that no considerations about a potential utilization of the code in a service need to be taken during the development.
The subsequent step of the development process is carried out by the \emph{service developer}, who wraps the existing forecasting or optimization code with the service framework in order to \emph{derive} a \emph{functional service}.
The latter is then \emph{operated} by the \emph{service provider} to make it usable for \glspl{EMS}.
It is worth noting that the service framework contains an operation concept (see Section \ref{sec_design_concept_operation_concept}) which supports service providers with their tasks.
The integration of the service into the \gls{EMS} is carried out by the \emph{EMS developer} who additionally needs to negotiate with the service provider with respect to the conditions under which the service can be used, including which data the \gls{EMS} must provide to the service.
The job of installation and maintenance of the \gls{EMS} is carried out by the \emph{EMS provider}, likely in close cooperation with the final \emph{user} of the system.

Returning to our running example, one might consider that the algorithm developer is an academic researcher who engineered the optimization algorithm for the \gls{BSS} within a project funded by an IT company that specializes in selling forecasting and optimization services.
The latter might act in the roles of the service developer and service provider, having several customers that specialized in designing and provisioning of \gls{EMS}.
The facility manager might then have ordered an \gls{EMS} for the building they supervise and therefore become the user of the EMS, thus also indirectly the user of the integrated services.

Finally, it is worth mentioning that we do consider, but not demand, that the aforementioned roles are distributed over institutions.
Nevertheless, it appears not unlikely in academic research projects that all roles are taken by a single institution, e.g. a research group, who might develop algorithms as well as an \gls{EMS} and test it in their own research facilities.
\section{Related Work}\label{sec_related_work}
This section analyses approaches related to ours from academia and industry.
\subsection{Frameworks for Service Development}
Most closely related to the present publication is \cite{maree_building_2022}, in which a service-based approach to create digital twins of buildings is presented.
Similar to our work, the paper strives to develop a framework.
The main difference is that their work is much broader, i.e. the framework covers not only forecasting and optimization but also data storage, thermal models of buildings, and how these can be learned from data.
Consequently, \cite{maree_building_2022} do not handle the aspect of forecasting and optimization services at a comparable depth as in the present work.
For example, it does not contain any requirements analysis, a detailed
discussion about the technical design and implementation of the services, nor does it provide a systematic approach to derive new forecasting or optimization services from existing code.
Furthermore, their work does not contain any hint about a potential publication of the corresponding source code. 
Thus we conclude that their work, unlike ours, is not a reasonable basis for deriving forecasting and optimization services for energy management applications.

On the other hand, larger \gls{ML} frameworks, like e.g. PyTorch\footnote{\href{https://pytorch.org/}{https://pytorch.org/}} MLflow\footnote{\href{https://mlflow.org/docs/latest/models.html}{https://mlflow.org/docs/latest/models.html}}, provide the functionality to expose an \gls{ML} model as web service with a \gls{REST} \gls{API}.
However, to the best of our knowledge, there is no solution that supports triggering the training models from \gls{API} calls out of the box\footnote{A detailed discussion why training models via \gls{API} calls is necessary is provided in Section \ref{sec_requirements_functional}.}, i.e. that allows fitting system-specific parameters as our framework does.
It thus appears that utilizing the framework developed in the present work is significantly more advantageous for service developers.
This is particularly the case if one considers that our framework has been explicitly designed to minimize the necessary effort for the development and operation of forecasting and optimization services for \glspl{EMS} at scale.

\subsection{Forecasting and Optimization Services}\label{sec_related_work_services}
Several publications have been identified, beyond \cite{maree_building_2022}, that utilize the concept of forecasting and/or optimization components wrapped into services \cite{Galenzowski2023,Lenk2020,hill_p2_2023,dengler_p4_2023,marinakis_big_2020,mohamed_service-oriented_2018}.
Regarding approaches not documented in academic publications one should first consider that several providers exist operating web \glspl{API} for the retrieval of weather-related data or forecasts, partly as free or commercial offering, e.g. Bright Sky\footnote{\href{https://brightsky.dev/}{https://brightsky.dev/}}, Open Meteo\footnote{\href{https://open-meteo.com/}{https://open-meteo.com/}}, SoDa\footnote{\href{https://www.soda-pro.com/}{https://www.soda-pro.com/}}, Solcast\footnote{\href{https://solcast.com/}{https://solcast.com/}} and Forecast.Solar\footnote{\href{https://forecast.solar/}{https://forecast.solar/}}.
The latter two offer additional services related to forecasting \gls{PV} power generation.
Closely related to the latter is NIXTLAs TimeGPT\footnote{\href{https://docs.nixtla.io/docs/getting-started-about\_timegpt}{https://docs.nixtla.io/docs/getting-started-about\_timegpt}}, a commercial service for generic time series forecasting.
Finally, it is worth mentioning the Building Energy Modeling\footnote{\href{https://exchange.se.com/develop/products/38969/building-energy-modeling-api}{https://exchange.se.com/develop/products/38969/building-energy-modeling-api}} service provided by Schneider Electric as part of their EcoStruxure platform, more details about the latter in the following section.
The service allows users to learn the thermal energy consumption pattern of buildings from data.

In contrast to our work, none of the publications or services referenced in this section present a framework for deriving forecasting and optimization services for energy management applications.
Thus, these services are neither in conflict with the present work nor do they provide any substantial input for the requirements analysis or design concept presented below.
While some of the mentioned offerings could be reasonably utilized by \glspl{EMS}, the main issue is that each of these services covers only a fraction of the typically required functionality, while none provides a generic approach to derive the remaining necessary forecasting and optimization services.
Nevertheless, the pure existence of these publications and services can be considered as strong advocacy for the general concept of forecasting and optimization services and, thus, the relevance of the present paper.
Furthermore, it is worth noting that the scientific research underlying the referenced publications would very likely have benefited substantially from using the framework proposed in the present work.
\subsection{Energy Management Systems, Platforms, Communities and Market Places}
A relevant platform that has been developed and utilized for years in EU projects is FIWARE~\cite{cirillo2019aStandard}. 
FIWARE is a general-purpose IoT Platform~\cite{cirillo2019aStandard} managed by the FIWARE Foundation~\cite{rodriguez2018fiware}.
It is used in various use cases, including smart farming~\cite{rodriguez2018fiware}, smart buildings, and smart grids~\cite{blechmann2023open}.
The heart of FIWARE is the so-called context broker, which receives data from data providers (e.g., sensors), stores the latest information, and provides it to data consumers (e.g., some service).
Aside from the context broker, there are various different solutions for data processing and storage that can be connected to the broker, as well as a set of ``smart data models'' that have been used in different applications.
With all these, the FIWARE ecosystem provides many different building blocks that can be used in energy management.
These building blocks, however, are very heterogeneous and generic, as their only shared foundation is the integration with the context broker and the underlying Next Generation Service Interface (NGSI).
Therefore, FIWARE is less an alternative to the proposed service framework and more a platform into which derived services could be integrated.

On the other hand, several commercial approaches exist that are similar to our service-based forecasting and optimization concept.
In particular noteworthy are the platform solutions from Siemens (Building X\footnote{\href{https://xcelerator.siemens.com/global/en/products/buildings/building-x.html}{https://xcelerator.siemens.com/global/en/products/buildings/building-x.html}}), Bosch (NEXOSPACE\footnote{\href{https://www.boschbuildingsolutions.com/xc/en/digital-services/}{https://www.boschbuildingsolutions.com/xc/en/digital-services/}}) and Schneider Electric (EcoStruxure\footnote{\href{https://www.se.com/ww/en/work/campaign/innovation/platform.jsp}{https://www.se.com/ww/en/work/campaign/innovation/platform.jsp}}).
The latter two appear conceptually similar, i.e. the platforms provide several functionalities for smart building operation, including energy management, but require that proprietary hardware (i.e. gateways) must be installed in the building that should be connected to the respective platform.
This is a clear contrast to the Siemens solution, which is advertised with an open \gls{API} concept and connectivity to third-party systems, while seemingly offering similar functionality like the other two.
No evidence was found that any of the three companies offers a framework like introduced in this work.
However, all three vendors claim that their platform solutions can be extended by third parties and offer a marketplace for applications that can be integrated.
However, publishing extensions on the marked places must be explicitly granted by the respective company and is subject to licensing fees.
Thus, we conclude that none of the three platforms is indeed a viable alternative to this work as neither empowers third parties to develop and operate forecasting and optimization services for energy management applications which are independent of the respective vendor.
Finally, it is worth mentioning that Schneider Electric and Bosch both offer \glspl{EMS} for private households, e.g. Bosch Smart Home\footnote{\href{https://www.bosch-smarthome.com/uk/en/}{https://www.bosch-smarthome.com/uk/en/}} and HEMSlogic\footnote{\href{https://shop.se.com/de/de/catalog/category/view/s/hemslogic/id/3252/}{https://shop.se.com/de/de/catalog/category/view/s/hemslogic/id/3252/}}.
While both of these systems apparently use some form of cloud-based optimization, there seems to be no possibility to directly interact with these forecasting and optimization services or to integrate third party services as an alternative.
However, we again perceive that the existence of the Siemens, Schneider Electric and Bosch solutions, providing cloud services for smart building operation and energy management, strongly advocates the concept underlying this work.

Finally, we find it important to discriminate our work from the Open Energy Platform\footnote{\href{https://openenergyplatform.org/}{https://openenergyplatform.org/}} as well as from OpenEMS\footnote{\href{https://openems.io/}{https://openems.io/}}, two projects well known among scientific researchers.
The first of these is a community effort to establish a collection of tools supporting the work with and publication of energy-related datasets, with a focus on energy system modeling.
This is clearly disjoint from our goal to provide tooling for the implementation of forecasting and optimization services for energy management applications.
On the other hand, OpenEMS is a fully functional, open source, and free to use \gls{EMS}.
While it does, in fact, contain a limited number of forecasting and optimization algorithms, providing these is not the essential task of the software.
The latter is particularly true as OpenEMS is usually operated on edge devices with little compute power, which limits the applicability of modern \gls{ML}-based forecasting and optimization methods.
However, it is absolutely reasonable to extend OpenEMS with an \gls{ESG} compatible client, to allow the integration of forecasting and optimization services derived with our framework, and we plan to demonstrate this in future work.
\section{Requirements Analysis}\label{sec_requirements}
Following the common procedure in software engineering, we begin with a systematic approach to assess the requirements that should be fulfilled by our service framework.
IEEE defines a requirement as 'A condition or capability that must be met or possessed by a system or system component to satisfy a contract, standard, specification, or other formally imposed documents' \cite{ieee_glossary_2002}.
It is worth noting that the traditional requirements engineering process, as defined in \cite{pohl_requirements_1996}, is tailored for the utilization in customer-specific software development.
In contrast, this work aims at developing a framework for a broad range of potential service developers, from academia and industry alike.
We thus employ a simple two-step process inspired from research on market-driven requirements engineering \cite{alves_study_2006,aurum_market-driven_2005}.
Hereby, the first step is an analysis of application areas.
To this end, typical applications of \glspl{EMS} in three different areas are described in Section \ref{sec_requirements_domain_analysis}.
Building upon this information, the requirements are documented in a semi-structured natural language specification \cite{washizaki_guide_2024}, using the following pattern: 'An <actor> must/should be able to <requirement>'.
Using this pattern, the requirements are documented in a transparent and consistent way that enables easy requirements verification.
Furthermore, we categorize the requirements into functional (Section \ref{sec_requirements_functional}) and non-functional (Section \ref{sec_requirements_non-functional}).
Here we follow Glinz's definition \cite{glinz_non-functional_2007}, where functional requirements describe a function a system must be able to perform, including component, behavioral, and functional aspects.
Quality and performance aspects, like throughput, reliability, and security, as well as constraining aspects, like physical or legal aspects, are summarized in non-functional requirements.
While the derivation of these requirements is generally based on the analysis provided in the following section, we are also guided by our broad experience in numerous projects in cooperation with relevant industry partners, where prototypical \glspl{EMS} for various scenarios have been developed and evaluated in large field tests, like e.g. MeRegio\footnote{\href{https://meregio.forschung.kit.edu/english/24.php}{https://meregio.forschung.kit.edu/english/24.php}}, C/sells\footnote{\href{https://www.wirsinteg.de/csells}{https://www.wirsinteg.de/csells}}, flexQgrid\footnote{\href{https://flexqgrid.de/english/}{https://flexqgrid.de/english/}}, and Smart East\footnote{\href{https://smart-east-ka.de/}{https://smart-east-ka.de/}}.
\subsection{Analysis of Application Areas}\label{sec_requirements_domain_analysis}
Basis for the requirement elicitation process is an analysis of application areas, with which we aim to provide more context and information on the environment, in addition to Section~\ref{sec_nomenclature}.
Domain knowledge and an understanding of the application's context is important for the quality of the requirements \cite{alebrahim2014structured, antonelli2012deriving}.
Therefore, to aid in the formulation of requirements, this analysis defines the relevant application areas or domains \cite{loucopoulos1988knowledge}, and is carried out by compiling typical applications for \glspl{EMS}, as well as their characteristics and distinguishing factors. 
For this purpose, we consider the three application areas private households, commercial buildings, as well as districts and areas separately, and concisely illustrate the individual goals and system specifics.
The latter, motivated by ISO 25010~\cite{iso25010}, is achieved by pointing out functional, efficiency, compatibility, interaction, reliability and safety, security, as well as maintainability and flexibility aspects.
Table \ref{table_domain_analysis} summarizes key differences and similarities between the application areas.


In private households, nearly two-thirds of the energy demand are used for space heating. 
They account for 27 \% of the final energy demand in the EU, of which only 25 \% is electricity \cite{eurostat_energy_2023}, but this is expected to rise due to the ongoing electrification of heat and transport due to electric vehicles and heat pumps \cite{ruhnau2019direct}.
Electricity generation by local \gls{PV} plants is also growing rapidly, leading to an increase in households that produce parts of their electricity consumption themselves (often called "prosumers") \cite{SOVACOOL2022112868, Kotilainen2019}.
The usual goal is to optimize local \gls{PV} usage and minimize the electricity needed from the public grid by using flexible devices, such as batteries, and shifting flexible electricity demand.
In order to do this, private prosumer households often use \glspl{EMS} to control their batteries, heat pumps, and/or charging processes \cite{zafar_hems_2020}.
They can work rule-based \cite{berkes2024sopevs}, which in simple cases also produces optimal results, for instance when there is a flat electricity tariff by storing all excess PV production and discharging whenever there is a deficit, or use all sorts of optimization algorithms, including mixed integer linear programming, genetic algorithms, particle swarm optimization and more \cite{srilakshmi2022energy, antunes2022comprehensive}.
Home \glspl{EMS} used in private households can either be provided as cloud services, e.g., by electricity providers, or operated locally on an edge device, e.g., a Raspberry Pi. 
In the latter option, \glspl{EMS} can be operated based on open-source smart home systems like Home Assistant\footnote{\href{https://www.home-assistant.io/}{https://www.home-assistant.io/}} or OpenHAB\footnote{\href{https://www.openhab.org/}{https://www.openhab.org/}} which can be installed and used by everyone, but may be limited in terms of computing power.
Hardware interoperability on the building level is a challenging task, due to a lack of standards and many vendor specific solutions.
The provided user interfaces vary depending on the intended user group.
Solutions like OpenHAB allow, for instance, the creation of own control rules, while others provide only simple visualizations.
All functions should be provided without interruption, to ensure user comfort, but usually outages would only lead to loss of comfort for the affected household(s).
Systems should be designed to avoid damage to devices and users.
Especially in private households, the limited computing power of local \glspl{EMS} makes it favorable to outsource optimization, load prediction, or \gls{PV} forecast to a cloud service. 
However, a strong argument for using local systems is the high level of privacy protection, as no data on electricity consumption, which can be used to draw conclusions about residents’ behavior, has to be shared with cloud providers \cite{boiko_edge-cloud_2024}.
Maintainability and (software) flexibility are crucial for \gls{EMS} developers and providers, especially for offering their customers continued safe and secure systems and allowing support for more and new hardware.

In commercial buildings, energy management algorithms like e.g. proposed by \cite{chen_gnu-rl_2019, ding_octopus_2019, oldewurtel_use_2012}, typically address the optimization of the \gls{HVAC} system, controlled centrally or for rooms individually.
Usually, these buildings are equipped with a \gls{BAS} on which a \gls{RBC} strategy is implemented.
The latter is replaced with an optimization-based approach given an \gls{EMS} is installed.
The efficiency, both from a software and energy perspective, varies with the employed algorithms and depends on the local systems \cite{al2021energy}.
Compatibility, like in the residential case, can be a challenge, however, with larger facilities and larger associated investments, customized integrations are more reasonable than in the residential case.
In commercial buildings, the correct operation can be of critical importance for the organization utilizing the building.
Therefore, the building optimization system might have to be executed on-premise to prevent outages caused by internet failures.
Using cloud-based energy management systems in commercial buildings can also come with challenges regarding privacy and security \cite{anthi_secure_2018}.
Commercial buildings might be utilized by organizations that are privacy-sensitive and thus do not permit data to be stored in the cloud. 
Other organizations, however, might be rather price-sensitive and hence prefer to use an optimization algorithm provided as cloud service while configuring the \gls{BAS} to fall back to \gls{RBC} in case of connection issues.
From a maintainer and vendor perspective, again, maintenance and flexibility are important for the operation of existing systems and the further development of the product.

\begin{table}
\caption{Comparison of three areas in which \glspl{EMS} are used.}
\label{table_domain_analysis}
\begin{tabular}{| p{2.5cm} | p{3.5cm} | p{3.5cm} | p{3.5cm} |}
\hline
   & \textbf{Private Households} & \textbf{Commercial Buildings}  & \textbf{Districts and Areas}  
\\ 
\hline
\textbf{Scale}  & Small  & Medium   & Large          
\\ 
\hline
\textbf{Exemplary applications} & Space heating, prosumers (\gls{PV}), heat pumps, electric vehicles & \gls{HVAC} systems, optionally \gls{PV} and electric vehicle charging stations & Buildings and energy grids, especially electrical distribution grids     
\\ 
\hline
\textbf{Optimization goals} & Maximize self-consumption, minimize energy demand from the grid and optimize with respect to electricity prices & Optimization of \gls{HVAC} systems to reduce energy demand while maintaining comfort & Cost-minimal operation of all generators and flexible loads and storages in the area 
\\ 
\hline
\textbf{Information used for optimization} & Building models, prediction of inflexible electricity demand and thermal energy demand, weather forecasts and local \gls{PV} supply & Weather forecast and occupancy predictions, if sensing devices available  & Non-standardized models due to higher level of aggregation and abstraction, current systems state, forecasting of energy-related time series 
\\ 
\hline
\textbf{Implementa-tion} & On-premise with local electricity-saving edge devices, or cloud-based services & On-premise, to prevent outages, based on existing \gls{BAS} with \gls{RBC} strategy & Dedicated server due to scale, can be on-premise for factories  
\\ 
\hline
\textbf{Privacy} & Privacy-sensitive data, e.g., residents’ behavior & Privacy-sensitive data possible – trade-off between privacy and price with cloud storage & Data aggregation can be used to protect privacy-sensitive information
\\ 
\hline
\end{tabular}
\end{table}
%

Districts and areas differ from those categories due to their size and, most importantly, the involvement of energy grids.
One major reason to conduct energy management on the level of facilities, districts, and even on a regional scale is grid operation.
With increasing decentralized generation, especially from renewable energy sources, and increasing demands from electrification, the need for monitoring the utilization of the grid and its power quality (see \cite{chawda2020comprehensive}) and actively influencing energy flows to prevent or resolve undesired situations is rising (e.g., \cite{volk2017grid}). 
Energy management on an area level, therefore, often considers the associated energy grids, especially in the case of micro-grids.
Another reason for area-level energy management is the optimal, e.g., cost-minimal, operation of all the generators, storage systems, and flexible loads in the area (e.g., \cite{roccotelli2022smart}).
Control of the different flexible devices and/or buildings in the area can be achieved with more or less direct mechanisms, ranging from direct device access to indirect, highly aggregated control signals \cite{foerderer2022automated}. 
The practical implementation and derived qualities like efficiency, reliability, and safety, depend on the selected orchestration mechanism, local regulation and the characteristics of the area in question, e.g., who owns the devices and energy grids and whether there are any fees for using the public grid in a given scenario.
Since, in a region, there can be any number of commercial and residential buildings combined, the interoperability challenge is amplified manifold.
Standardized interfaces, models, and \glspl{EMS} for each building can alleviate this challenge \cite{khalid2024smart}.
Users may be provided with user interfaces for checking the current regional status and history, or making inputs, such as electric vehicle charging settings.
In districts and areas, privacy-sensitive data, e.g. of many households, may need to be protected, which can be done using data aggregation due to the larger scale \cite{kursawe_privacy_2011, varenhorst2024enhancing, langer2013privacy}.
Here, a higher level of aggregation and abstraction is additionally beneficial to keep the amount of data that has to be managed and the computation times on an acceptable level, also resulting in a need for different models.
Aggregation is especially important on higher grid and system levels. 
Smart energy-optimized areas may, for instance, aggregate their flexibility on the feeder level (e.g., \cite{volk2017grid}). 
In such a scenario, control signals need to be disaggregated for their implementation upon reception.
Reliable and safe operation are especially important on the regional level, as faults may leave many buildings without energy.
For achieving reliable and safe operation, maintainability and flexibility in software are especially helpful on this level, compared to the other two.

From a general perspective, the basic building blocks needed for energy management in all three application areas
are very similar, that is, functionality for determining and assessing the current systems state, forecasting of energy-related time series, optimization of load schedules or similar control signals, and controllers implementing the schedules.
The implementation, however, may vary due to the different properties and specific demands present in the application areas.
In all three areas, privacy concerns need to be taken into account due to the presence of privacy-sensitive information.
%
%
\subsection{Functional Requirements}\label{sec_requirements_functional}
With the analysis of application areas presented in the previous section and the described typical energy management applications in mind, we now derive functional requirements.

The first requirement directly results from the service and stakeholder concept discussed in Section \ref{sec_nomenclature}. It is:
\req{FR01}{A service developer must be able to derive a functional service with the service framework from existing forecasting or optimization code.}

Here, the intention of the service provider clearly is to allow \glspl{EMS} to utilize the existing forecasting or optimization algorithm by interacting with the web \gls{API} of the service\footnote{Note that we formulate the remaining requirements about the derived service to improve readability. Later, in particular in Section \ref{sec_design_concept_service_components}, we discuss that the requirements need to be fulfilled by the service framework.}.
Hence the second and third requirements are:
\req{FR02}{An \gls{EMS} must be able to interact with the service over a web \gls{API} provided by the service.}
\req{FR03}{An \gls{EMS} must be able to \emph{request} a forecast or optimized schedule utilizing the API of the service.}
 
The forecasting or optimization algorithms wrapped by the service framework will usually require some form of input data.
Considering a \gls{PV} power generation forecast as example, this could be the global position and time of the target system.
Furthermore, the data format returned by a forecasting or optimization algorithm will obviously be specific to it and should contain all the information the algorithm needs for processing the expected result.
The latter includes constraints that should be obeyed by optimization algorithms.
Therefore, the fourth requirement is:
\req{FR04}{A service developer must be able to specify the format of the input and output data exchanged due to an \gls{EMS} request for a forecast or optimized schedule from the service \gls{API}.}

Some services may implement \emph{system-specific parameters} that must be \emph{fitted} utilizing historical measurements of the system subject to forecast or optimization as a prerequisite for high-quality results.
Note that the algorithm for fitting the system-specific parameters is considered to be a part of the existing forecasting or optimization code.
As services should be usable by a large number of \glspl{EMS}, this fitting process should be manageable via the service \gls{API}\footnote{Note that web services exposing machine learning models usually implement a different approach, i.e. that the training will be carried out before deployment and hence that the parameters of the model are identical for all clients. In order to emphasize this difference, we explicitly refer to \emph{fitting system-specific parameters} instead of \emph{training models}. Furthermore, it is worth noting that our proposed approach will also work for foundation model based forecasting approaches, like e.g. TimeGPT introduced in Section \ref{sec_related_work_services}. While such algorithms may or may not need to fit system-specific parameters for good performance, exposing them as services is nevertheless reasonable to enable a widespread application in \glspl{EMS}.}.
Returning to the \gls{PV} power generation forecast example, one could conceive that the forecasting code contains a small neural network that has been trained using the power generation of several \gls{PV} systems, thus representing an average system.
However, if power generation measurements of a specific \gls{PV} system are available, it is possible to adapt (fit) the weights and bias terms of the neural network (\emph{system-specific parameters}) such that the prediction error is minimized for the specific system.
The fitting procedure is part of the service and the fitting process can be initiated by calling the respective \gls{API} endpoint with the required input data, which would be a time series of historic power generation data for the \gls{PV} power generation forecast example.
Furthermore, we need to consider privacy-sensitive users, i.e. users that do not accept any of their data to be stored in a cloud database, which implies that it must be possible for \glspl{EMS} to store the fitted parameters locally\footnote{One could argue that a user who opposes storing data in a cloud database will likely not want to use an \gls{EMS} that utilizes forecasting or optimization services operated by an external service provider at all. On the other hand, service providers might guarantee that data exchanged with a service is deleted immediately after processing, which seems like a fair compromise between privacy and cost efficiency.}.
Finally, we need to consider \glspl{EMS} that are not capable of reliably storing historic recordings of measurements or fitted parameters on-premise, e.g. very likely a large fraction of \glspl{EMS} operating in private households.
While this seems like a major difference at first glance, it turns out that this scenario imposes no additional requirements for the service.
The discussion behind this finding is out of scope at this point but can be found in Appendix \ref{appendix_interaction_patterns}.
The resulting requirements are thus:
\req{FR05}{An \gls{EMS} must be able to fit system-specific parameters of a service utilizing its \gls{API}.}
\req{FR06}{A service developer must be able to specify the format of the input and output data exchanged while an \gls{EMS} interacts with the \gls{API} of a service to fit the system-specific parameters.}
\req{FR07}{An \gls{EMS} must have the option to store fitted system-specific parameters locally.}

\gls{API} calls made by an \gls{EMS} may take a significant amount of time before the result becomes available.
Consider e.g. the service providing \gls{PV} power generation forecasts used as a running example for which fitting the system-specific parameters involves training a neural network that might take several minutes to hours.
On the other hand, computing forecasts or optimized schedules might require a decent amount of time too, e.g. if computing an optimized schedule for a larger building involves solving a complex linear program.
As such response times are different from typical values of web services, we formulate it as an additional requirement:
\req{FR08}{An \gls{EMS} must be able to make calls to the \gls{API} of the service which may take several hours to compute.}

Finally, it is well known that \emph{documentation} is important for the widespread adoption of \glspl{API} \cite{hunter_irresistible_2017,jin_designing_2018}.
Here documentation refers to the description of the functionality of a service, in particular its \gls{API} and the data format for interactions with the latter.
Furthermore, development efforts can be reduced by automatically generating the documentation from the corresponding source code, which is additionally beneficial as it prevents that changes in the code are not reflected in the documentation.
The resulting final functional requirement is thus:
\req{FR09}{A service developer should be able to automatically generate a documentation for the \gls{API} of a service.}
\subsection{Non-Functional Requirements}\label{sec_requirements_non-functional}
Extending the content above, this section presents non-functional requirements for the service framework.
To this end, we first consider the envisioned target state that professional service providers operate services which are utilized by a large number of \glspl{EMS}.
Thus, the availability of these services is very likely of critical importance for the intended functioning of a large number of \glspl{EMS}.
This implies that service providers need to apply state-of-the-art computing cluster techniques for operation.
Furthermore, the data exchanged between \gls{EMS} and services may contain sensitive information and should thus be encrypted\footnote{Note that it is additionally reasonable to demand that the data exchanged between \gls{EMS} and service cannot be altered in transit. However, we do not add this point as a separate requirement as it is automatically fulfilled if the communication is securely end-to-end encrypted, e.g. with \acrshort{HTTPS}}, especially as a large share of \glspl{EMS} will likely communicate with services over the public internet.
Finally, operating services may require significant compute resources and energy.
A service provider may, hence, wish to restrict access to services to certain \glspl{EMS}.
This leads to the following requirements:
\req{NFR01}{A service provider must be able to operate services with high availability and scalability.}
\req{NFR02}{An \gls{EMS} must be able to communicate with the service over an encrypted connection.}
\req{NFR03}{A service provider must be able to restrict access to a service to authorized \glspl{EMS}.}

As the correct functioning of the proposed service framework is substantial for the stable operation of the derived services, it becomes clear that service developers and providers must be convinced that the framework is implemented correctly to adopt it.
Furthermore, service providers will likely not utilize the framework, if no reasonable maintenance concept exists, which suggests that future problems in the framework will be addressed and solved quickly.
\req{NFR04}{A service developer/provider should be able to validate the correct implementation of the service framework.}
\req{NFR05}{A service developer/provider should be able to verify that the service framework is actively maintained.}

Beyond the commercial aspect, an important intended application of the service framework is to empower academic researchers to derive functional services from existing forecasting or optimization code.
For this, one needs to consider the limited resources typical for academic research, which implies that the task of deriving a service should require minimal effort\footnote{This is obviously beneficial for service providers with a commercial background too.}.
Furthermore, it should be regarded that some service developers might not have a strong expertise in applied informatics but should still be able to utilize the proposed framework.
An example to illustrate this demand could be a project with public funding dedicated to energy management in private households carried out by a consortium of research groups.
In such a case, it may appear beneficial to integrate a research group dedicated to energy meteorology to develop a forecast service for \gls{PV} power generation without demanding that this group cares about the operation and implementation details of the service.
This leads to the following requirements:
\req{NFR06}{A service developer should be able to derive a service with minimal effort from an existing forecasting or optimization algorithm.}
\req{NFR07}{A service provider should be able to operate a service without requiring expert knowledge about IT infrastructure.}

A service operating for a longer time may need continuous development work by both service developer and provider, e.g. in order to maintain or even improve performance and usability.
Such efforts could include breaking changes, like e.g. an adaption of the data format which is not backward compatible.
In order to give \gls{EMS} developers time to adjust to those changes, it is common to operate an old and a new version in parallel.
However, this implies that the \gls{EMS} developer must be able to select which version of a service should be utilized, thus leading to the following requirement:
\req{NFR08}{An \gls{EMS} developer must be able to specify which version of a service should be utilized.}

While deriving \reqref{FR09} above, we have argued that documentation is important for the adoption of \glspl{API} by \gls{EMS} developers.
However, beyond the pure existence of a documentation, it appears reasonable to demand that the latter should allow \gls{EMS} developers to rapidly comprehend the \gls{API} of a service.
Furthermore, the main intention of a \gls{EMS} developer reading the documentation is likely to implement a client in order to interact with the \gls{API} of a service.
For this, we demand minimal effort of implementation again, assuming it will likely support widespread adoption of the corresponding service.
Hence, our final two requirements are:
\req{NFR09}{An \gls{EMS} developer should be able to quickly understand the \gls{API} of a service by utilizing the documentation.}
\req{NFR10}{An \gls{EMS} developer should be able to implement a client to interact with the \gls{API} of a service with minimal effort.}

We are convinced that the requirements derived in the present and previous section are a solid foundation for deriving a framework for provisioning forecasting and optimization algorithms as web services for \glspl{EMS}, and demonstrate this suitability below.
\section{Design Concept}\label{sec_design_concept}
Based on the requirements discussed above, we introduce the design concept of our proposed service framework in this section.
To this end, we first present the \gls{API} design, in particular as providing an \gls{API} for forecasting and optimization code is the core functionality of our proposed solution.
Based on the \gls{API} design, we proceed to describe the internal operation of a service derived from the framework and finally conclude this section with a discussion about service operation.
\subsection{API Design}\label{sec_design_concept_api_design}
As a first step, it is necessary to choose the paradigm on which the \gls{API} of our proposed service framework should be based.
We consider well-established approaches for web-based \glspl{API} (\reqref{FR02}). These are \gls{REST} \cite{fielding_architectural_2000}, \gls{RPC} (in particular gRPC\footnote{\href{https://grpc.io/}{https://grpc.io/}}) as well as GraphQL\footnote{\href{https://graphql.org/}{https://graphql.org/}}.
As all three candidates are generally suited to satisfy the functional requirements we focus on the non-functional requirements in order to select the best suited paradigm.
The relevant requirements are understandability (\reqref{NFR09}) as well as ease of client implementation (\reqref{NFR10}).
It is generally perceived that \gls{REST} is the most favorable approach regarding these demands \cite{hunter_irresistible_2017,jin_designing_2018}, which is therefore selected.

As a next step, we define the functionality of the \gls{API} that is provided by the service framework.
The selection of \gls{REST} implies that all communication between client and service will use the \gls{HTTP} and that the functionality must be mapped to \glspl{URL}.
It should be noted that we will only note down the relative part of \glspl{URL} for the sake of brevity and to highlight that the domain is not relevant for the structure of the \gls{API}, i.e. we use \texttt{/endpoint1/} instead of the full notation \texttt{https://some-service.example.com/endpoint1/}.
The selection of \gls{REST} furthermore implies that we need to define which \gls{HTTP} method (like e.g. \texttt{GET}, \texttt{POST}, \texttt{PUT}, or \texttt{DELETE}) must be used in order to receive a desired outcome while interacting with a specific \gls{URL}.
We will henceforth refer to the combination of \gls{HTTP} method and (relative) \gls{URL} as \emph{\gls{API} method}.
Further introduction about web-based communication over \gls{HTTP} can be found in the usual introductory texts or as a short summary in \cite{jin_designing_2018}.

\reqref{FR03} dictates that \glspl{EMS} must be able to retrieve a forecast or optimized schedule from a service.
Furthermore, we need to consider that a service may take minutes or even hours to compute the result (\reqref{FR08}).
As especially the latter is far beyond typical timeouts of \gls{HTTP} servers\footnote{The default timeout of nginx for a read operation is, for example, 60 seconds.} it is infeasible to directly return the computation result.
Instead, we define three \gls{API} methods to overcome this issue:
\\\ttinnewline{POST\quad/\string{version\string}/request/}
\ttinnewline{GET  \quad/\string{version\string}/request/\string{task\_ID\string}/status/}
\ttinnewline{GET  \quad/\string{version\string}/request/\string{task\_ID\string}/result/}
The intended interaction of an \gls{EMS} with these \gls{API} methods is as follows:
\begin{enumerate}
	\item The \gls{EMS} issues a \texttt{POST} call to the \texttt{/\string{version\string}/request/} endpoint containing the required input data (see \reqref{FR04}).
	Note that \texttt{\string{version\string}} is a placeholder that must be filled with the desired version of the targeted service, which is required to satisfy \reqref{NFR08}, and could e.g. have a value of \texttt{v2}, see the example provided below.
	The service checks whether the input data is correct.
	If that is the case the service starts computing the result in the background and returns an ID, e.g. \texttt{123}, associated to this task (more details about the latter are provided in the following two sections).
	\item Using the ID of the request, the \gls{EMS} should issue calls to the endpoint: \texttt{GET /\string{version\string}/request/\string{task\_ID\string}/status/}\footnote{Note that this polling mechanism seems a bit inelegant at first glance. However, alternatives have severe downsides too, e.g. WebHooks imply the need for the \gls{EMS} to be exposed on the network while WebSockets don't integrate well into documentation and tooling of \gls{REST} \glspl{API}. See Chapter 2 in \cite{jin_designing_2018} for a more detailed discussion.}.
	Note that \texttt{\string{task\_ID\string}} is a placeholder too.
	Regarding the example above, the endpoint would be \texttt{/v2/request/123/status/}.
	For each call, the service will compute and return the status of the computation, that is one of \texttt{queued}, \texttt{running}, or \texttt{ready}\footnote{Note that we have not added a \texttt{failed} state as \texttt{ready} just implies that the result endpoint can be called. The information whether the requested computation has succeeded or failed is provided by the \gls{HTTP} status code returned while calling the \texttt{/\string{version\string}/request/\string{task\_ID\string}/result/} endpoint. This is the standard approach for communication over \gls{HTTP}.}.
	\item Once the service has finished processing the result and a \texttt{ready} status has been observed, the \gls{EMS} can issue a \texttt{GET} call to the \texttt{/\string{version\string}/request/\string{task\_ID\string}/result/} endpoint to retrieve the output of the computation, i.e. the forecast or optimized schedule.
\end{enumerate}
Additionally, to the ones defined above, it is necessary to specify \gls{API} methods to allow fitting system-specific parameters of a service in order to satisfy \reqref{FR05}.
As potentially long processing times (\reqref{FR08}), as well as versioning (\reqref{NFR08}), need to be considered again, it appears reasonable to take over the concept introduced above and define the respective \gls{API} methods as:
\\\ttinnewline{POST\quad/\string{version\string}/fit-parameters/}
\ttinnewline{GET  \quad/\string{version\string}/fit-parameters/\string{task\_ID\string}/status/}
\ttinnewline{GET  \quad/\string{version\string}/fit-parameters/\string{task\_ID\string}/result/}
The intended usage of the \texttt{/fit-parameters/} endpoints is equivalent to the pattern discussed for \texttt{/request/} above.

Finally, and in order to satisfy \reqref{NFR03}, we need to consider authorization, which implies authentication, to allow service providers to restrict access to specific clients.
Following \cite{spath_rest_2023,saeed_security_2022,kornienko_principles_2021} the currently best practice for \gls{REST} \glspl{API} is token-based\footnote{A token is considered as a string that is sent with every \gls{HTTP} request to the webserver (the service in our case), most commonly in the \gls{HTTP} header.
The token is usually specific for each client and allows the webserver to validate if the request is permitted or not, e.g. by looking up the permissions associated with the particular token in a database.}
authentication, in particular the utilization of \glspl{JWT}.
\glspl{JWT}, as defined in \cite{jones_json_2015}, are special tokens that can be cryptographically validated.
Regarding the scope of this work, \glspl{JWT} are issued by a dedicated identity provider (see Section \ref{sec_design_concept_operation_concept}) to the client software.
The signature of the token allows the service to check locally\footnote{Locally means here that the token can be validated without the lookup operation mentioned in the previous footnote, which supports scalability.} whether a request of a client should be granted or not.
Further details about the application of \glspl{JWT} for web security are given in \cite{saeed_security_2022}.
%
%
%
\subsection{Service Components}\label{sec_design_concept_service_components}
The concept described in this section arises from the requirements \reqref{FR01} and \reqref{NFR06}, i.e. that it should be possible to derive a functional service from an existing forecasting or optimization code with minimal effort.
Especially the latter (\reqref{NFR06}) imposes that as much functionality as possible should be realized by the service framework in order to keep the implementation effort for the service developer low.
\begin{figure}
\centering
\FIG{\includegraphics[width=0.7\columnwidth]{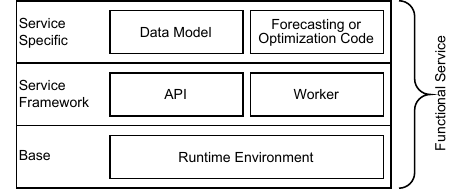}}{\caption{Components of a service derived with the service framework}\label{fig_service_components}}
\end{figure}
To this end we define a functional service to consist of several components that can be grouped into three categories as summarized in \figref{fig_service_components}:
\begin{enumerate}
	\item Base: Containing the components necessary for executing the code of the service.
	\item Service Framework: Containing all components generic to all services.
	\item Service Specific: Containing all components a service provider must implement to derive a functional service.
\end{enumerate}

The primary component of the service framework is the \gls{API} as defined above.
Additionally, we need to consider that computing the result of a request (as well as fitting system-specific parameters) might take significant time (\reqref{FR08}), while the \gls{REST} \gls{API} of the service should respond immediately.
Hence, it is necessary to decouple the \gls{API} from the interaction with forecasting or optimization code.
Therefore, we introduce the \emph{worker}, which is a second component provided by the service framework that is executed in a dedicated process and that is responsible for computing the requested results.
It is worth noting that this concurrent processing is crucially important, as, otherwise, executing a forecasting or optimization code might block the \gls{API} from responding to other calls from clients\footnote{
Decoupling the \gls{API} from computing results furthermore allows more sophisticated load management, like queuing tasks until resources for computation are available, while still allowing the \gls{API} to be responsive to client calls.}.

The service specific category contains the actual payload of the service, i.e. the forecasting or optimization code.
Additionally, the requirements \reqref{FR04} and \reqref{FR06} need to be considered, i.e. that service developers must be able to specify the format of the input data for calls to \texttt{/request/} and \texttt{/fit-parameters/} as well as the output format returned by the corresponding \texttt{/result/} \gls{API} methods.
We will refer to the part of the implementation that defines these formats as \emph{data model}\footnote{We chose \emph{data model} to be consistent with the terminology used by the framework employed for implementing the \gls{API} component (i.e. FastAPI, see Section \ref{sec_implementation_api} below) which uses \emph{model}. However, the latter collides with a potential model used in the forecasting or optimization code. Hence, we use data model.}.
At this point, we will not further specify possible characteristics of the data model as the latter are tightly connected to the implementation of the \gls{API}.
However, we will proceed discussing this topic in Section \ref{sec_implementation_api}.
\subsection{Service Architecture}\label{sec_design_concept_service_architecture}
It was discussed in the previous section (\ref{sec_design_concept_service_components}) that a functional service must contain an \gls{API} as well as a worker component and that these should be operated in distinct processes for performance reasons.
It should be noted at this point that the service specific components, i.e. the data model as well as the forecasting or optimization code, are perceived to be parts of the \gls{API} and worker components.
The service specific components are consequently not explicitly mentioned in this section to promote readability.

Following from the execution of service components in distinct processes the necessity arises to establish some form of inter-process communication to allow services to operate.
Considering the simplest case, i.e. a service consisting of two processes, one for the \gls{API} and one for the worker, communication between the two could be established quite simply using queues.
However, we need to consider \reqref{NFR01}, i.e. that service providers should be able to operate services with high availability and scalability.
From the latter follows that services might consist of multiple instances of \gls{API} and worker components, while high availability imposes that the service components might be distributed over several machines.
The usual approach in such a scenario, which is utilized in this work, too, is to leverage a \emph{message broker} for communication between components.
\begin{figure}
\centering
\FIG{\includegraphics[width=0.7\columnwidth]{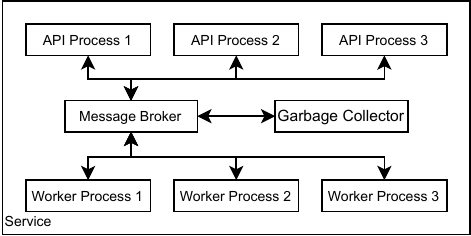}}{\caption{Internal architecture of a service with processes of components and communication between processes}\label{fig_service_architecture_full}}
\end{figure}
The resulting communication pattern within a service would be as follows:
\begin{itemize}
	\item Every valid \texttt{POST} call to a \texttt{/request/} or \texttt{/fit-parameters/} endpoint should lead to the creation of a \emph{task}, an object carrying the necessary information for computing the result, by the \gls{API}. The latter should assign an ID to the task and publish the task on the message broker to invoke the workers. Finally, the \gls{API} should return the task ID to the client.
	\item A worker process should fetch the task from the message broker and start computing the result by invoking the forecasting or optimization code. Additionally, the worker should regularly publish status updates on the processing progress on the broker.
	\item In case of a call to a \texttt{/status/} endpoint, the \gls{API} should fetch the latest status information about the corresponding task from the broker and return this information to the client.
	\item If a \texttt{/result/} endpoint is called, the \gls{API} should fetch the result from the broker and return the result to the client.
\end{itemize}
Finally, it must be considered that a result that is fetched from the broker should not be deleted on the broker immediately.
For example, consider that the communication between client and \gls{API} might be interrupted, in which case the client will likely retry to fetch the result.
However, in order to prevent unlimitedly growing memory consumption of the message broker, it is required to operate an additional component, i.e. the \emph{garbage collector}.
The duty of the latter is to delete the task-related data from the broker that are likely not to be required anymore.

The resulting internal architecture of a service, including communication channels, is indicated in Figure \ref{fig_service_architecture_full}.
It should be appreciated that the service architecture introduced above is strictly designed to support scalability and high availability, in particular, as subsequent calls from clients are not required to hit the same \gls{API} process again. That is, it is not required that a call to the \texttt{/status/} endpoint will be handled by the same \gls{API} process that created the corresponding task, as all task-related information is shared on the broker.
Furthermore, it is possible to scale the number of \gls{API} as well as worker processes to the actual load induced by clients.
\subsection{Operation Concept}\label{sec_design_concept_operation_concept}
Building up on our considerations introduced above, we will conclude the service design in this section by discussing the missing piece, which is the operation concept.
To this end, it should be recalled first that (in this work) a service consists of several components (like \gls{API} or worker) and that in reality operation might require that several instances of each component are executed in parallel, possibly distributed over several machines (see Section \ref{sec_design_concept_service_architecture}).
We have referred to the latter as processes in order to distinguish the running instance from the implementation.
The management of these processes, commonly referred to as orchestration, is a tedious activity best left to specialized and well-established applications.
In consequence, the first step towards the operation concept is to select an appropriate orchestration software, where it needs to be considered that service providers have different demands here, ranging from cloud computing professionals of commercial companies with a primary interest in high availability and scalability (\reqref{NFR01}) to academic researchers with no experience with the utilization of orchestration software (\reqref{NFR07}).
Systematic comparisons of orchestration applications \cite{malviya_comparative_2022,jawarneh_container_2019,yang_comparison_2019} suggest Kubernetes\footnote{\href{https://kubernetes.io/}{https://kubernetes.io/}} for the first group and Docker Swarm\footnote{\href{https://docs.docker.com/engine/swarm/}{https://docs.docker.com/engine/swarm/}} for the second.
Both orchestrators require that the processes of the service are wrapped in containers.
As Docker\footnote{\href{https://www.docker.com/resources/what-container/}{https://www.docker.com/resources/what-container/}} containers are supported by both they are selected.

Beyond the execution of the service components as described in the previous sections, provisioning of functional services to \glspl{EMS} requires that a service provider operates additional supportive applications.
The need for the first one of these supportive applications arises from the simple demand that the \gls{API} containers must be accessible for the \glspl{EMS} (\reqref{FR04} and \reqref{FR06}), as well as that requests from clients should be distributed over the \gls{API} containers of a service, in order to allow highly available and scalable operation (\reqref{NFR01}).
We will refer to the application that fulfills this duty as \emph{gateway}.
However, it should be noted that such applications are often alternatively named \emph{reverse proxy} or \emph{ingress}\footnote{\href{https://kubernetes.io/docs/concepts/services-networking/ingress/}{https://kubernetes.io/docs/concepts/services-networking/ingress/}}, the latter in particular in the context of Kubernetes.
It is common, too, that the gateway encrypts the communication with the client, i.e. by utilizing the \gls{HTTPS}, on behalf of the payload application, in our case, the service.
This procedure removes the burden of integrating the security-relevant and vastly complex encryption logic into the code base of the \gls{API} component while still satisfying the requirement that communication between client and service must be secure (\reqref{NFR02}).
The second supportive application is necessary to issue \gls{JWT} tokens to clients in order to satisfy \reqref{NFR03} as discussed in Section \ref{sec_design_concept_api_design}.
Such applications are usually referred to as \gls{IdP} and the current best practice choice for issuing \glspl{JWT} is the \gls{OIDC} protocol \cite{scott_best_2021,kornienko_principles_2021}.
Hence, our operation concept follows this best practice and intends that a well-established \gls{IdP} software, like e.g. Keycloak\footnote{\href{https://www.keycloak.org/}{https://www.keycloak.org/}}, is used in order to administrate access control and to issue \gls{JWT} tokens to the client software using \gls{OIDC}.
Finally, it is worth noting that it may not be necessary that every service provider operates an \gls{IdP} by themselves.
Especially in the context of academic research, it is likely sufficient that one partner operates an \gls{IdP}, e.g. for a project, while services can still be distributed over several partners.
This significantly reduces the effort for providing services for those partners that do not serve the \gls{IdP}, as operating the latter is a complex task requiring specialized knowledge.
This possibility should support especially academic researchers with backgrounds other than informatics.
%
\section{Implementation}\label{sec_implementation}
In this section, we present the reference implementation of the design concept introduced above, which is released alongside this paper as an open source repository and which we refer to as \acrfull{ESG}.
The latter is implemented in Python, which has been selected due to the availability of many relevant software libraries for the implementation of forecasting and optimization algorithms, like e.g. PyTorch\footnote{\href{https://pytorch.org/}{https://pytorch.org/}}, TensorFlow\footnote{\href{https://www.tensorflow.org/}{https://www.tensorflow.org/}} or Pyomo\footnote{\href{https://www.pyomo.org/}{https://www.pyomo.org/}}.
Furthermore, Python offers high connectivity to other programming languages, especially the integration of C and C++ is natively supported\footnote{\href{https://docs.python.org/3/extending/extending.html}{https://docs.python.org/3/extending/extending.html}}.
Furthermore, the latter allows the integration of code written in other programming languages that can be compiled to C, e.g. go\footnote{\href{https://pkg.go.dev/cmd/cgo}{https://pkg.go.dev/cmd/cgo}}.
While there may be no simple solution to integrate code written in some programming languages, like e.g. especially Java, into the \gls{ESG} framework, the concepts described in Section \ref{sec_design_concept} are sufficiently generic to allow an implementation of a similar framework in other programming languages.
However, this is not subject to this work, which presents only one implementation of the design concept, which is the \gls{ESG} framework written in Python. 

It is worth noting that the following sections are, on purpose, rather concise, i.e. cover the relevant points to demonstrate that the proposed implementation satisfies the respective requirements.
Additional details will be omitted for the sake of conciseness, in particular as the implementation should be continuously developed further as an open source project (see Section \ref{sec_community_concept}), which will likely lead to significant differences compared to the code state at the publication date of this paper.
However, further information is provided in the \gls{ESG} repository\footnote{\href{https://github.com/fzi-forschungszentrum-informatik/energy-service-generics/}{https://github.com/fzi-forschungszentrum-informatik/energy-service-generics/}}, including examples how services and clients can be implemented.
\subsection{Worker, Garbage Collector, and Inter-Process Communication}\label{sec_implementation_worker}
In order to ensure that our implementation is robust and to minimize future maintenance effort, we refrain from developing a custom approach for the inter-process communication between \gls{API} and worker.
Instead, we utilize Celery\footnote{\href{https://docs.celeryq.dev/}{https://docs.celeryq.dev/}}, a well-established, stable, and open-source Python library for distributed task execution.
It is worth noting that Celery is particularly well suited for use cases similar to ours, i.e. to decouple client-facing web components from worker processes in a scalable fashion.
Furthermore, Celery supports multiple message brokers, including the well-established Redis\footnote{\href{https://redis.io/}{https://redis.io/}} and RabbitMQ\footnote{\href{https://rabbitmq.com/}{https://rabbitmq.com/}}, which gives service providers flexibility to choose a broker matching their demands, tech stack, and/or skill set of employees. 
We provide an implementation of a generic worker, which invokes the service specific forecasting or optimization code in the \gls{ESG} package.
The worker makes use of Celery to interact with the message broker and implements the functionality described in Section \ref{sec_design_concept_service_architecture}.
Finally, Celery provides a garbage collector functionality.
Dependent on the choice of message broker the latter may be available without the need for a dedicated process, more details are provided in the respective part of the documentation\footnote{\href{https://docs.celeryq.dev/en/stable/userguide/configuration.html\#result-expires}{https://docs.celeryq.dev/en/stable/userguide/configuration.html\#result-expires}}.
\subsection{API}\label{sec_implementation_api}
The first step towards implementing the \gls{API} is to select an appropriate framework to build upon.
Considering the choice of Python as programming language, it follows that two well-established web frameworks are available\footnote{Django is a well-established and Python-based web framework too. However, Django is primarily intended for the development of websites and heavily relies on communication with a database, and is hence not regarded a suitable candidate here.}, these are Flask\footnote{\href{https://palletsprojects.com/p/flask/}{https://palletsprojects.com/p/flask/}} and FastAPI\footnote{\href{https://fastapi.tiangolo.com/}{https://fastapi.tiangolo.com/}}.
Several requirements need to be considered in order to select one of the two candidates objectively.
In particular, these are \reqref{FR09} (service developers should be able to automatically generate a documentation), \reqref{NFR09} (\gls{EMS} developers should be able to quickly understand the \gls{API} using the documentation) as well as \reqref{NFR10} (\gls{EMS} developers should be able to implement a client with little effort).
We have selected FastAPI because it is generally considered to have a better integration of the OpenAPI specification\footnote{\href{https://swagger.io/specification/}{https://swagger.io/specification/}} and tool ecosystem \cite{singh_flask_2023,krebs_developing_2022}.
OpenAPI schema is a gold standard for documenting \gls{REST} \glspl{API}.
FastAPI is capable of automatically generating and hosting OpenAPI schema documents as well as serving Swagger UI\footnote{\href{https://swagger.io/tools/swagger-ui/}{https://swagger.io/tools/swagger-ui/}}.
The latter is an interactive \gls{API} documentation building up on the former, thus satisfying \reqref{FR09}.
Furthermore, an interactive documentation is considered to be especially helpful for client developers to quickly understand an \gls{API} \cite{hunter_irresistible_2017}, which makes FastAPI a reasonable choice to fulfill \reqref{NFR09}.
Another helpful tool from the OpenAPI ecosystem is Swagger Codegen\footnote{\href{https://swagger.io/tools/swagger-codegen/}{https://swagger.io/tools/swagger-codegen/}} which allows the automatic generation of large shares of client code for an impressive number of programming languages, thus satisfying \reqref{NFR10}.

Using FastAPI, our implementation of the \gls{API} component is capable of serving the endpoints defined in Section \ref{sec_design_concept_api_design}.
It is fully functional, apart from the definition of the data models, i.e. the input data for \texttt{POST /request/} and \texttt{POST /fit-parameters/} as well as the corresponding \texttt{GET /result/} \gls{API} methods, which need to be defined by the service provider.
It is worth noting that the definition of the data models is simple and fast to implement in order to fulfill \reqref{NFR06}.
A practical example and additional details are provided in Section \ref{sec_evaluation_data_model}.
Furthermore, our implementation of the \gls{API} component puts the communication concept defined in Section \ref{sec_design_concept_service_components} into action, i.e. it transposes the \gls{HTTP} calls of the clients into interactions with the message broker using the Celery framework introduced above, but only after the corresponding \gls{JWT} of the call has been verified and checked.
We utilize the PyJWT\footnote{\href{https://pyjwt.readthedocs.io/en/stable/}{https://pyjwt.readthedocs.io/en/stable/}} package for the latter.
Finally, it is worth mentioning that our implementation solely utilizes the \gls{JSON} data format for all data exchange, as the latter is interpretable for humans and machines alike and furthermore considered to be the best choice for \gls{REST} \glspl{API} \cite{hunter_irresistible_2017}.
\subsection{Other Functionality}
Besides the implementation of the \gls{API} and the process management system, the \gls{ESG} package provides additional useful functionality for the implementation of and interaction with services.
Particularly relevant is a generic client that can be used to trigger calls to services from Python source code.
Furthermore, the package contains building blocks for data models, in order to reduce the effort for implementing these as well as to prevent code redundancy, see Section \ref{sec_evaluation_data_model} for further details.
Additionally, useful utility functions are available, like e.g. to parse pandas\footnote{A popular Python library for analysis and manipulation of time series data.} "DataFrames" from \gls{JSON} data.
%
%
\section{Community Concept}\label{sec_community_concept}
Above, we have discussed the design and implementation of the service framework that satisfies the requirements presented in Section \ref{sec_requirements_functional} and \ref{sec_requirements_non-functional},
with two exceptions: So far, we have not addressed that service developers and providers should be able to verify the correct implementation (\reqref{NFR04}) and that the service framework should be actively maintained (\reqref{NFR05}).
We satisfy the first of these two requirements by releasing our implementation (i.e. the \gls{ESG} package) as free and open-source repository, including an extensive documentation.
A link to the repository is provided in the section "\nameref{sec_supplementary_material}" at the end of this paper.
In order to fulfill \reqref{NFR05}, we strive to find a concept that ensures the continuous maintenance of the service framework.
To that end, we first consider that typical reasons for modern open source projects to fail are usurpation by competitors, the project being not useful anymore or the lack of time and interest of developers~\cite{coelho2017WhyModernOpen}.
While we certainly cannot prevent that forecasting and optimization services may be not useful anymore at some point in the future, we strive to anticipate the remaining reasons leading to the abandoning of open source projects by initiating the \emph{Open Energy Services} community.
Following the classification and discussion about possible organizational forms of communities that maintain open source projects in \cite{eckert2019AloneTogetherInterorganizational}, we chose the form of an autonomous community, i.e. one that is not closely tied to nor owned by a company or organization, as we want to avert that service providers may object utilizing the framework due to interest clashes with the former.
Furthermore, we have considered current best practices \cite{fogel_producing_2022,mateos-garcia_institutions_2008} while setting up the internal structures of the community in order to let decentralized and vital processes for exchange, development, and maintenance flourish.

The efforts of our community are not to be limited to the maintenance of the framework.
Instead, the goal is to directly support the widespread adoption of \gls{EMS} by connecting service developers, service providers and \gls{EMS} developers, all of which are invited to become members of the community.
We strive to develop a number of forecasting and optimization services and release these as open source repositories.
Furthermore, it is intended to operate a selection of services for the public\footnote{Note that we plan to provide these services free of charge. However, we will likely take measures to limit the load to our compute resources, e.g. by restricting access to registered clients or by limiting the number of \gls{API} calls allowed per time.}.
Here, the concept is to first provide basic building blocks for energy optimization, like e.g. forecasting services for electric loads or \gls{PV} power generation, in order to bootstrap energy management-related research and development activities, particularly by further academic institutions.
The main communication medium is the community website\footnote{\blinded{\href{https://open-energy-services.org/}{https://open-energy-services.org/}}{<URL blinded for review>}}, where we will release further information about recent activities and developments arising from our continuous efforts.
Furthermore, and inspired by the FAIR principles for research software \cite{lamprecht_towards_2020}, the website contains a registry for service-related open source repositories, as well as a registry for operated services.
The main intention is that the community website should become the primary source of objective information about forecasting and optimization services for energy management applications in the future.

The founding members of the Open Energy Services community are the research institutions to which the authors of this paper belong.
However, we explicitly invite other institutions and persons to join in order to establish a solid basis for the development and operation of forecasting and optimization services that enable energy management solutions at scale.
\section{Evaluation}\label{sec_evaluation}
The goal of this paper is to support the widespread adoption of \glspl{EMS} in order to unlock the flexibility and energy savings potentials of end consumers.
To this end, we contribute a concept for a software framework that allows the derivation of fully functional services from existing forecasting or optimization code with ease.
Furthermore, we publish an open-source implementation of our proposed approach.
Additionally, we establish a community in order to ensure the future maintenance of the framework, but also to support the widespread adoption of forecasting and optimization services in energy management applications.

We demonstrate in this section that our contributions are actually sufficient to reach our claimed goal of this paper, i.e. that our framework and community concept are useful for establishing forecasting and optimization services for energy management applications.
To this end we begin by showcasing that our framework allows service developers to derive forecasting or optimization services with ease by providing a practical example for the implementation of a service.
As it is also intended that the derived services can be integrated effortlessly into \glspl{EMS}, we demonstrate the latter by providing two practical examples of client scripts that allow interacting with the previously derived example service.
We continue our practical demonstration by exploring the scalability of services in order to verify that the services derived with our framework are indeed capable of serving thousands of \glspl{EMS}.
We conclude our evaluation with a structured comparison of the derived requirements with our concept and implementation, to illustrate the completeness of our work.
\subsection{Deriving Services}\label{sec_evaluation_example}
As a first step towards proving the relevance of this work, we demonstrate how a simple (but fully functional) \gls{PV} power generation forecast service can be implemented using the proposed framework.
The following subsections present and discuss the implementation of the components of the service.
It should be noted that the forecasting algorithm of the illustrated service, including the quality of the forecasts the service could compute, is not of particular interest.
Instead, the aim of this section is to demonstrate the value of the framework, which can be perceived by realizing how short the code listings below are.
Besides the forecasting or optimization algorithm, which must be implemented anyway, deriving the functional example service requires less than 100 lines of code.
On the other hand, at the time of writing, the \gls{ESG} repository contained about 2400 lines of code, excluding examples and documentation.
It is thus obvious that utilization of the \gls{ESG} framework is much more work-efficient compared to the implementation of services from scratch, as the latter would require the developer to implement a larger fraction of the code provided by the framework.
In fact, implementing services from scratch might be far more time-consuming than the proportion of lines of code suggests.
This is, in particular, the case as the code using the framework is quite simple, as demonstrated below, while some parts provided by the framework are complex and need to be implemented carefully, e.g. due to security concerns or impacts on runtime behavior.

It is worth noting that the code listings provided below are part of the \gls{ESG} repository\footnote{\blinded{\href{https://github.com/fzi-forschungszentrum-informatik/energy-service-generics/tree/main/docs/examples/basic\_example}{https://github.com/fzi-forschungszentrum-informatik/energy-service-generics/tree/main/docs/examples/ basic\_example}}{<Link removed for blind review>}}, whereby the online version might be adapted to any potential future changes of the \gls{ESG} framework.
We thus suggest all those wishing to reproduce this example, i.e. run the example service, to use the files provided in the repository.
\subsubsection{The Forecasting or Optimization Code}
The forecasting or optimization code is the payload of the service, i.e. the goal of applying the \gls{ESG} framework is to make this component accessible to \glspl{EMS}.
The \gls{ESG} framework has been designed to make integration of existing forecasting or optimization code simple, which is demonstrated in the example by utilizing the popular pvlib\footnote{\href{https://github.com/pvlib/pvlib-python}{https://github.com/pvlib/pvlib-python}} for computing the PV power production forecast.
However, it is worth noting that the framework does not induce any restrictions on the forecasting or optimization code wrapped by it.
For example, linear programs, classical statistical models, or fully black-box machine learning approaches are all possible.
Even model ensembles can be realized, either as a single service or as a service that calls other services as ensemble members.

In order to allow pvlib to compute the forecasts, it is necessary to provide the corresponding input data to the library.
The first part of this input data is the specification of the PV system for which the forecast should be computed.
We assume that the PV system is sufficiently described by the geographic position, i.e. latitude and longitude, as well as the geometry of the PV system, i.e. azimuth and inclination, and the peak power.
All other options to describe the PV system offered by pvlib\footnote{\href{https://pvlib-python.readthedocs.io/en/stable/user\_guide/modelchain.html}{https://pvlib-python.readthedocs.io/en/stable/user\_guide/modelchain.html}} are neglected to make this example not more complex than necessary.
The second part of the required input to compute \gls{PV} power prediction consists of meteorological forecast data, especially forecasts of solar irradiance.

Before we proceed with the discussion about the handling of input data by the \gls{ESG} framework, it is worth recalling that the framework supports two types of endpoints, these are \texttt{/request/} and \texttt{/fit-parameters/}. This differentiation arises from the requirement \reqref{FR05}, i.e. that the \gls{ESG} framework should allow forecasting or optimization code that uses \gls{ML} approaches, which induces that some parameters of a model must be fitted utilizing observations of the target system, see Section \ref{sec_design_concept_api_design} for details.
Regarding the example above, we assume that the service is intended to produce \gls{PV} power generation forecasts for systems for which geometry and peak power values may be unknown and need to be estimated from power production measurements.
The parameter fitting has been implemented with a simple least squares approach, although it should be noted that this choice has no particular relevance for the present example.
Thus, the input data necessary to obtain a forecast is separated into two groups: latitude and longitude are \emph{arguments} while azimuth, inclination, and peak power are \emph{parameters}.
It is worth noting that parameters are not necessarily interpretable as in this example, e.g. weights and bias terms of a neural network could be parameters too.
Finally, it should be considered that it may not be reasonable to demand all input data as client input.
In the present example, the service fetches the meteorological data automatically from a third-party web service, which would, in practice, make the interaction with the service more convenient and less error-prone for the client.
Finally, the following points concerning the code listing below should be noted:
\begin{enumerate}
	\item The format of \lstkw{input_data} and \lstkw{output_data} is implicitly defined in the corresponding data models, which are introduced in the following section.
	\item The functions \lstkw{predict_pv_power}, \lstkw{fetch_meteo_data} as well as \lstkw{fit_with_least_squares} have been omitted from the listing, as the practical implementation details of those are not of particular interest for the scope of this work. However, the code of the omitted functions can be found in the repository of the \gls{ESG} framework.
	\item Implementing \lstkw{fit_parameters} is optional and can be omitted for services without fittable parameters. An example without fittable parameters could be a service wrapping the AutoPV algorithm proposed in \cite{meisenbacher_autopv_2023}.
\end{enumerate}
%
\begin{lstlisting}[language=python,linerange={24-25,215-999},caption={Integration of the forecasting or optimization code.}]
from esg.utils.pandas import series_from_value_message_list
from esg.utils.pandas import value_message_list_from_series

def handle_request(input_data):
    arguments = input_data.arguments
    parameters = input_data.parameters
    meteo_data = fetch_meteo_data(
        lat=arguments.geographic_position.latitude,
        lon=arguments.geographic_position.longitude,
    )
    pv_power = predict_pv_power(
        lat=arguments.geographic_position.latitude,
        lon=arguments.geographic_position.longitude,
        azimuth=parameters.pv_system.azimuth_angle,
        inclination=parameters.pv_system.inclination_angle,
        peak_power=parameters.pv_system.nominal_power,
        meteo_data=meteo_data,
    )
    output_data = {"power_prediction": value_message_list_from_series(pv_power)}
    return output_data

def fit_parameters(input_data):
    arguments = input_data.arguments
    measured_power = series_from_value_message_list(input_data.observations.measured_power)
    meteo_data = fetch_meteo_data(
        lat=arguments.geographic_position.latitude,
        lon=arguments.geographic_position.longitude,
        past_days=90,
    )
    measured_power = series_from_value_message_list(input_data.observations.measured_power)
    fitted_pv_system = fit_with_least_squares(
        lat=arguments.geographic_position.latitude,
        lon=arguments.geographic_position.longitude,
        meteo_data=meteo_data,
        measured_power=measured_power,
    )
    return fitted_pv_system
\end{lstlisting}
It should be noted that the two functions defined above, i.e. \lstkw{handle_request} and \lstkw{fit_parameters}, are the only part of the implementation of the service that actually interacts with the forecasting or optimization algorithm.
The duty of these functions is to correctly invoke the forecasting or optimization code with the necessary input data.
\subsubsection{The Data Model}\label{sec_evaluation_data_model}
In addition to the forecasting or optimization code introduced above, the data model is the second component that is service specific and which must thus be defined by the service developer.
The data models define the format of the data the client exchanges with the service.
For a service without fittable parameters, i.e. a service with \texttt{/request/} endpoints only, it is sufficient to define the arguments required for computing the request as well as the result of the computation.
The corresponding data models are called \lstkw{RequestArguments} and \lstkw{RequestOutput}.
It is worth noting that \lstkw{RequestArguments} could also contain constraints that optimization services should account for.
Furthermore, \lstkw{RequestArguments} will likely often contain a field that defines the temporal resolution of the generated forecast or optimized schedule.
Alternatively, some services may implement a fixed temporal resolution that cannot be manipulated by the client.
In the present example, the \gls{PV} power generation forecast is always returned on an interval of 15 minutes.

In the case of a service with fittable parameters, it is additionally necessary to define the data format for the input and output data for the \texttt{/fit-parameters/} endpoints.
The data models specifying the input for the fitting process are referred to as \lstkw{FitParameterArguments} and \lstkw{Observations}, and the corresponding output is \lstkw{FittedParameters}.
As the simple \gls{PV} power generation forecast service used as an example is designed to provide functionality to fit parameters, it is necessary to define all five data models introduced above.
The corresponding implementation is shown in Listing \ref{lst_service_data_model}.
\begin{lstlisting}[language=python,linerange={22-999},label={lst_service_data_model},caption={Definition of the data models.}]
from esg.models.base import _BaseModel
from esg.models.datapoint import ValueMessageList
from esg.models.metadata import GeographicPosition, PVSystem
from pydantic import Field

class RequestArguments(_BaseModel):
    geographic_position: GeographicPosition

class RequestOutput(_BaseModel):
    power_prediction: ValueMessageList = Field(description="Prediction of power production in W")

class FitParameterArguments(_BaseModel):
    geographic_position: GeographicPosition

class Observations(_BaseModel):
    measured_power: ValueMessageList = Field(description="Measured power production in W")

class FittedParameters(_BaseModel):
    pv_system: PVSystem
\end{lstlisting}
At this point, it is worth noting that the \gls{ESG} package provides ready-to-use building blocks for data models.
For example, in the code above, \lstkw{GeographicPosition} is imported from \gls{ESG}.
The former is a data model too, which defines that a geographic position consists of latitude and longitude.
Furthermore, it is crucial to note that the data models serve additional functionality beyond the definition of the data format.
Particularly important is the provisioning of documentation of the format in human-readable form (e.g. the \lstkw{description} in the example above) as well as defining permitted ranges for values, the latter is utilized by the derived service to automatically validate the input data provided by clients.
An example for such a permitted range could be to enforce that values for latitude must be in the range of $-90^\circ \dots 90^\circ$. This rule is, in fact, implemented in \lstkw{GeographicPosition}, although this is not directly visible in the code example above.
However, it can be perceived by inspecting the source code of the \lstkw{GeographicPosition} data model\footnote{\blinded{\href{https://github.com/fzi-forschungszentrum-informatik/energy-service-generics/blob/main/source/esg/models/metadata.py}{https://github.com/fzi-forschungszentrum-informatik/energy-service-generics/blob/main/source/esg/models/ metadata.py}}{<Link removed for blind review>}}.

\subsubsection{The Worker}
The worker component is responsible for executing the tasks, i.e. computing requests or fitting parameters by invoking the forecasting or optimization code, as well as task scheduling.
Further details are provided in Sections \ref{sec_design_concept_service_components} and \ref{sec_design_concept_service_architecture}.
The \gls{ESG} framework utilizes the Celery library for implementing the worker, but extends the latter with functionality to make the implementation of services more convenient, for example by utilizing the data models for de-/serialization of input and output data.
Thus, the main objective for implementing a worker is to wire up the data models with the forecasting or optimization code, which should usually require a rather simple program, as displayed in Listing \ref{lst_service_worker}, for the \gls{PV} power generation forecast example service.
\begin{lstlisting}[language=python,linerange={22-999},label={lst_service_worker},caption={Definition of the worker tasks.}]
from esg.service.worker import celery_app_from_environ
from esg.service.worker import invoke_fit_parameters
from esg.service.worker import invoke_handle_request

from data_model import RequestArguments, RequestOutput
from data_model import FittedParameters, Observations
from data_model import FitParameterArguments
from fooc import fit_parameters, handle_request

app = celery_app_from_environ()

@app.task
def request_task(input_data_json):
    return invoke_handle_request(
        input_data_json=input_data_json,
        RequestArguments=RequestArguments,
        FittedParameters=FittedParameters,
        handle_request_function=handle_request,
        RequestOutput=RequestOutput,
    )

@app.task
def fit_parameters_task(input_data_json):
    return invoke_fit_parameters(
        input_data_json=input_data_json,
        FitParameterArguments=FitParameterArguments,
        Observations=Observations,
        fit_parameters_function=fit_parameters,
        FittedParameters=FittedParameters,
    )
\end{lstlisting}
\subsubsection{The API}
The \gls{API} component connects the worker with the client by allowing the latter to trigger the computation of requests or fitting of parameters as well as retrieving the corresponding results.
To this end, the \gls{API} component has to check the client input for validity and create the computation tasks.
Furthermore, the \gls{API} component handles authentication and authorization of clients.
More details about the \gls{API} design are provided in Section \ref{sec_design_concept_api_design}.

The implementation of the \gls{API} component is available ready-to-use in the \gls{ESG} framework.
However, in order to operate the \gls{API} it is necessary, similar to the worker, to wire up the \gls{API} with the other components, in particular with the data model and the worker.
Furthermore, some information like name and version number must be provided too.
Nevertheless, the necessary code to instantiate an \gls{API} component is trivially simple and shown in Listing \ref{lst_service_api}.
\begin{lstlisting}[language=python,linerange={22-999},label={lst_service_api},caption={Instantiation of the \gls{API} component.}]
from data_model import RequestArguments, RequestOutput
from data_model import FittedParameters, Observations
from data_model import FitParameterArguments
from worker import request_task, fit_parameters_task

api = API(
    RequestArguments=RequestArguments,
    RequestOutput=RequestOutput,
    FittedParameters=FittedParameters,
    Observations=Observations,
    FitParameterArguments=FitParameterArguments,
    request_task=request_task,
    fit_parameters_task=fit_parameters_task,
    title="PV Power Prediction Example Service",
)

if __name__ == "__main__":
    api.run()
\end{lstlisting}
\subsubsection{The Service}
Following the operation concept for services, as given in Section \ref{sec_design_concept_operation_concept}, the last remaining step for the service developer to derive functional services is to build docker images that can be run, e.g. on Kubernetes.
It is necessary to build two distinct images, one for the \gls{API} (which includes the data model) and one for the worker (which includes the data model and the forecasting or optimization code).
The build instructions for both images are implemented as Dockerfile\footnote{\href{https://docs.docker.com/reference/dockerfile/}{https://docs.docker.com/reference/dockerfile/}}, see Listings \ref{lst_service_docker_api} and \ref{lst_service_docker_worker}.
\begin{lstlisting}[language=docker,label={lst_service_docker_api},caption={Dockerfile for the \gls{API} container.}]
FROM energy-service-generics:latest-service

COPY api.py data_model.py worker.py /source/service/

CMD ["/source/service/api.py"]
\end{lstlisting}
\begin{lstlisting}[language=docker,label={lst_service_docker_worker},caption={Dockerfile for the worker container.}]
FROM energy-service-generics:latest-service-pandas

RUN pip install pvlib scipy
COPY data_model.py fooc.py worker.py /source/service/

ENTRYPOINT [ "celery" ]
CMD [ "--app", "worker", "worker", "--loglevel=INFO" ]
\end{lstlisting}
\subsection{Client Implementation}\label{sec_evaluation_ease_of_client}
Above, see \reqref{NFR10}, we have argued that a key requirement for the widespread integration of services into \glspl{EMS} is the ease of client implementation.
In order to demonstrate the latter we provide an example for a minimal client, implemented as shell script, capable of retrieving \gls{PV} power generation forecasts from the service introduced in the previous section in Listing \ref{lst_client_minimal}.
It should be noted that the example assumes that the service is reachable via the \texttt{localhost} network address, i.e. that the client is executed on the same machine as the service.
Note further that the latest version of the code listed below, i.e. the version adapted to potential future changes of the framework,  is provided as part of the \gls{ESG} repository\footnote{\href{https://github.com/fzi-forschungszentrum-informatik/energy-service-generics/tree/main/docs/examples/minimial\_client}{https://github.com/fzi-forschungszentrum-informatik/energy-service-generics/tree/main/docs/examples/ minimial\_client}}.
\begin{lstlisting}[language=bash,linerange={18-999},label={lst_client_minimal},caption={Minimal example of a client script that allows retrieving \gls{PV} power generation forecasts from the example service.}]
SERVICE_BASE_URL=${SERVICE_BASE_URL:-http://localhost:8800}
SERVICE_VERSION="test_version"

# Request a PV power generation forecast from the basic example service.
response=$(
    curl -X 'POST' \
    "${SERVICE_BASE_URL}/${SERVICE_VERSION}/request/" \
    -d '{
        "arguments": {
            "geographic_position": {
                "latitude": 49.01365,
                "longitude": 8.40444
            }
        },
        "parameters": {
            "pv_system": {
                "azimuth_angle": 0,
                "inclination_angle": 30,
                "nominal_power": 15
            }
        }
    }'
)

# Extract the ID of the task from the JSON response.
task_ID=$(echo $response | jq -r .task_ID)

# Poll status endpoint until status is ready.
status="unknown"
while [ $status != "ready" ]
do
    sleep 1
    response=$(
        curl -X 'GET' \
        "${SERVICE_BASE_URL}/${SERVICE_VERSION}/request/${task_ID}/status/"
    )
    status=$(echo $response | jq -r .status_text)
done

# Fetch and print the result.
response=$(
    curl -X 'GET' \
    "${SERVICE_BASE_URL}/${SERVICE_VERSION}/request/${task_ID}/result/"
)
echo $response | jq
\end{lstlisting}
One can perceive from inspecting the code above that the client logic is indeed very simple, thus easy to implement.
The script is written in standard Unix Shell syntax and uses only two additional packages, \texttt{curl} for making \gls{HTTP} requests and \texttt{jq} for parsing \gls{JSON}.
The script follows the \gls{API} concept derived in Section \ref{sec_design_concept_api_design}.
That is, a request is created with the first \texttt{curl} call.
The large nested structure underneath defines the input arguments and parameters that are provided to the service.
More discussion about the latter, including an explanation about the fields, is provided in Section \ref{sec_evaluation_data_model}.
Next, the script extracts the task ID of the created request and polls the \texttt{/status/} endpoint until the task has reached the \texttt{ready} status and finally fetches the result.
Note that the interaction with the \texttt{/fit-parameters/} endpoint of the service follows the same pattern and is thus omitted here for brevity.

While the service integration used in a production \gls{EMS} would likely require more functionality, e.g. to parse the result into the format the \gls{EMS} expects or to handle network errors, the concept of the client always remains the same.
In fact, the \gls{ESG} package contains a ready-to-use generic client\footnote{\href{https://github.com/fzi-forschungszentrum-informatik/energy-service-generics/blob/main/source/esg/clients/service.py}{https://github.com/fzi-forschungszentrum-informatik/energy-service-generics/blob/main/source/esg/clients/ service.py}} which provides this additional functionality and reduces the implementation effort further.
\begin{lstlisting}[language=python,linerange={20-999},label={lst_client_esg},caption={Example of a client script that allows retrieving \gls{PV} power generation forecasts from the example service using the generic client provided in the \gls{ESG} package.}]
import os

from esg.clients.service import GenericServiceClient
from esg.models.base import _BaseModel
from esg.models.datapoint import ValueMessageList
from esg.models.metadata import GeographicPosition, PVSystem
from esg.service.worker import compute_request_input_model
from pydantic import Field

SERVICE_BASE_URL = os.getenv("SERVICE_BASE_URL")

class RequestArguments(_BaseModel):
    geographic_position: GeographicPosition

class FittedParameters(_BaseModel):
    pv_system: PVSystem

RequestInput = compute_request_input_model(
    RequestArguments=RequestArguments, FittedParameters=FittedParameters
)

class RequestOutput(_BaseModel):
    power_prediction: ValueMessageList = Field(description="Prediction of power production in W")

client = GenericServiceClient(
    base_url=SERVICE_BASE_URL, InputModel=RequestInput, OutputModel=RequestOutput
)

client.post_obj(
    input_data_obj={
        "arguments": {"geographic_position": {"latitude": 49.01365, "longitude": 8.40444}},
        "parameters": {
            "pv_system": {"azimuth_angle": 0, "inclination_angle": 30, "nominal_power": 15}
        },
    }
)

print(client.get_results_obj())
\end{lstlisting}
The code in Listing \ref{lst_client_esg} demonstrates that the generic client class, which is provided as part of the \gls{ESG} package, reduces the effort to the implementation of the data models as well as configuring the endpoint of the service the client should use.
It is worth noting that specifying the data models has been left on purpose for the developer of the client, although it is technically possible that the client fetches the latter automatically from the service.
However, the idea is that the implementation of the data model on the client side documents the data structure that the downstream application, i.e. the \gls{EMS}, is designed for.
That is, it allows the client application to detect any changes in the format of the data provided by the service.
While such changes should not happen in theory, see discussion in \reqref{NFR08} and about versioning in Section \ref{sec_design_concept_api_design}, they might still occur due to mistakes made by developers of services.
Here, it is likely much simpler to debug an error that is thrown directly in the client code than some error deep downstream in the application using the data, which might have no obvious direct connection to the service and its data format.

Finally, we would like to point out that the generic client provided by the \gls{ESG} package can only be used in applications capable of executing Python programs.
Any developer working on an application not capable of the latter will likely need to implement the client logic from scratch.
However, the effort for such an activity should be rather limited as the generic client has been implemented in less than 400 lines of code.
As an alternative, the developer could resort to partly automatically generate the client program using Swagger Codegen, see Section \ref{sec_implementation_api} for further details.
\subsection{Scalability of Services}
One of the core claims of this work is that the presented framework enables operation of forecasting and optimization services for potentially thousands of \glspl{EMS}.
In order to demonstrate that this claim is indeed legitimate, we present an experiment that assesses the scalability in the present section.
\subsubsection{Experiment Design}
The goal of the experiment design described here is to create a situation in which we can examine the influence of horizontal scaling on the operation of a service using the proposed framework.
In particular, the experiment is intended to investigate the scalability related to communication overhead between worker and \gls{API} instances.
The latter is of particular importance as, given sufficient available funds, compute resources can be bought in nearly infinite amounts, which means that the actual limiting factor to scalability is the inter-process communication, i.e. in our case the communication between \gls{API} and worker containers as well as the intermediary message broker.

In contrast, very limited computing resources have been available for this experiment.
In particular, the experiment was carried out on a single virtualized server with access to 64GiB of main memory and 12 cores of an Intel Xeon 4116 \acrshort{CPU}.
We used Ubuntu 22.04\footnote{\href{https://ubuntu.com/}{https://ubuntu.com/}} as operating system and Microk8s\footnote{\href{https://microk8s.io/}{https://microk8s.io/}}, a Kubernetes distribution especially suitable for research and development, as container orchestration engine.
A dedicated service, henceforth referred to as \emph{scalability tester service}, has been implemented using the \gls{ESG} framework for the sake of this experiment.
The implementation consists of the previously described components, i.e. data model, worker, \gls{API}, as well as forecasting or optimization code.
Especial consideration has been devoted to designing the latter as, given the limited resources and desired high number of tasks, it is obviously not possible to conduct any operation which requires a non-neglectable amount of compute resources.
On the other hand, the component representing the forecasting or optimization code should provoke a realistic call pattern by the clients.
That is, the clients will usually poll the \texttt{/status/} endpoint several times before the result becomes available.
We expect this polling mechanism to have a significant impact on the communication load inside a service, as every status call triggers a lookup operation on the message broker.
In order to satisfy both demands, we have implemented a simple sleep operation of ten seconds representing the forecasting or optimization code, as it does not induce \acrshort{CPU} load while still blocking the worker and preventing immediately available results.
Furthermore, a code simulating clients has been implemented.
The latter uses the 100 instances\footnote{It has not been possible to use a single client for each call as this exhausted the network resources, i.e. the number of dynamic ports, on the machine executing the respective script.} of the generic client (as introduced in Section \ref{sec_evaluation_ease_of_client}) that issue in total 10.000 requests in very close time proximity.
Utilizing the standard settings for the generic client, each client polls the \texttt{/status/} endpoint once per second and retrieves the results as they become ready.
It is worth noting that each worker instance is configured to spawn 1000 threads, i.e. is capable of processing 1000 tasks in parallel. 
From this follows that there is an ideal bound for the processing time of the requests that is dependent on the number of worker instances.
For example, neglecting all communication overhead, two workers of the scalability tester service are capable of processing 2.000 requests (because of the 1.000 threads per worker) every 10 seconds (due to the sleep time of 10 seconds representing the forecasting or optimization code), thus leading to a minimum compute time of 50 seconds for all 10.000 requests. 
Further details about the implementation of the scalability tester service and the corresponding clients are omitted here for brevity.
However, the source code is provided in the \gls{ESG} repository\footnote{\href{https://github.com/fzi-forschungszentrum-informatik/energy-service-generics/tree/main/docs/examples/scalability\_example}{https://github.com/fzi-forschungszentrum-informatik/energy-service-generics/tree/main/docs/examples/ scalability\_example}}.

\subsubsection{Experiment Execution and Results}
In order to execute the experiment, the scalability tester service was deployed to the single-node Kubernetes instance introduced above.
Utilizing the latter, the \gls{API} and worker instances were scaled to a selected number, henceforth referred to as \emph{replication factor}, i.e. it was taken care that this many instances of each the worker and \gls{API} container were executed.
The \gls{API} and worker containers were using a single-node Redis\footnote{\href{https://redis.io/}{https://redis.io/}} instance as message broker, a popular choice given the implementation of the \gls{ESG} package\footnote{\href{https://docs.celeryq.dev/en/stable/getting-started/backends-and-brokers/index.html\#redis}{https://docs.celeryq.dev/en/stable/getting-started/backends-and-brokers/index.html\#redis}}.
The actual experiment was conducted by executing the code invoking the clients.
The latter was run on the personal laptop of one of the authors that was connected over the public internet to the service for extended realism.
\begin{figure}
\centering
\FIG{\includegraphics[width=0.8\columnwidth]{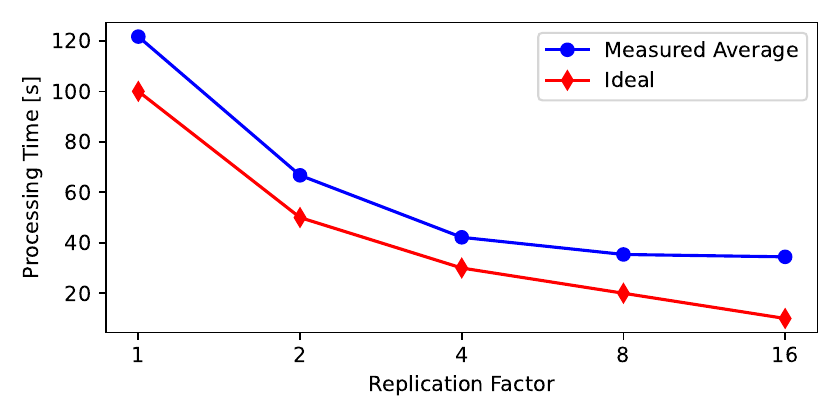}}{\caption{Measured and ideal time for processing 10.000 requests over replication factor}\label{fig_scalability_results}}
\end{figure}

The result, as shown in Figure \ref{fig_scalability_results}, is the average elapsed time starting directly before the program issued the first request and ending after all results have been available and transferred back.
The measurements were repeated three times and the difference between the minimum and the maximum of the measured times were for all replication factors one second or less, the results appear thus robust and reliable.  
It can be seen that the time required for processing 10.000 requests shrinks from initially ca. 120 to approx. 35 seconds as the replication factor is increased from 1 over 2, 4, 8, and finally 16.
It is thus clearly possible to mitigate a growing number of requests by replicating worker and \gls{API} instances as the total number of requests per time unit is reciprocal to the processing time required for handling a fixed number of requests.

An interesting observation from the inspection of the measured results is that increasing the replication factor from 8 to 16 yields only a rather small gain of one second in real processing time, while the ideal processing times suggest a significant reduction from 20 to 10 seconds.
Analysis of the \acrshort{CPU} and memory consumption of the machine on which the experiment was executed on yielded no evidence that exhausted hardware could have been the reason.
However, it appears likely that this behavior has been caused by the high number of threads (16.000 with the largest replication factor!), which very likely imposes a significant overhead for the operating system to switch tasks between these.
In the latter case distributing the worker instances over more than one machine should improve the performance even further.
While, at the time of writing, we do not possess the technical resources to validate this claim experimentally, we plan to catch up on this matter once we have access to a more sophisticated and better equipped Kubernetes cluster.

Finally, it is worth mentioning that the processing time of 35 seconds for 10.000 requests is roughly equivalent to 250.000 \glspl{EMS} issuing one request per 15 minutes, which appears a solid foundation for forecasting and optimization services at scale.
However, we estimate that using more sophisticated computing resources, it should rather easily be possible to serve several millions of \glspl{EMS}.
%
%
\subsection{Comparison of Requirements with Concept and Implementation}
As last part of our evaluation, this section presents a systematic comparison of the requirements derived in this work with the realization of our proposed framework and community concept.
This comparison is provided in Table \ref{table_functional_requirements} for the functional and non-functional requirements defined in Section \ref{sec_requirements_functional} and  \ref{sec_requirements_non-functional}.
One can easily perceive that our contributions, i.e. design and implementation of the framework as well as the community concept, do actually satisfy all requirements by inspecting the provided table.

As the requirements have been carefully derived from the current state of the art of energy management applications, it is concluded that our proposed framework and community concept are indeed valuable tools for the implementation and widespread distribution of forecasting and optimization services.

\newpage
\begin{longtable}{| l | p{4cm} | p{8cm} |}
\caption{Comparison of functional and non-functional requirements with actual realization in service framework and community concept.}
\label{table_functional_requirements}\\
\hline
\textbf{ID} & \textbf{Description} & \textbf{Realization}
\\
\hline
\textbf{FR01} &
A service developer must be able to derive a functional service with the service framework from existing forecasting or optimization code. &
A service developer can derive a functional service by extending the generic components provided by the service framework with service specific components. See Section \ref{sec_design_concept_service_components} for details.
\\
\hline
\textbf{FR02} &
An \gls{EMS} must be able to interact with the service over a web \gls{API} provided by the service. &
An \gls{EMS} can interact with a service (that utilizes the service framework) over a \gls{REST} \gls{API}. Further details can be found in Sections \ref{sec_design_concept_api_design} (design) and \ref{sec_implementation_api} (implementation).
\\
\hline
\textbf{FR03} &
An \gls{EMS} must be able to \emph{request} a forecast or optimized schedule utilizing the API of the service. &
An \gls{EMS} can request a forecast or optimized schedule from a service (that utilizes the service framework) by interacting with the \texttt{/request/} endpoints defined in Section \ref{sec_design_concept_api_design}.
\\
\hline
\textbf{FR04} &
A service developer must be able to specify the format of the input and output data exchanged due to an \gls{EMS} request for a forecast or optimized schedule from the service \gls{API}. &
A service developer can specify the format of the input and output data for the \texttt{/request/} endpoints by defining the data model. The concept is defined in Section \ref{sec_design_concept_service_components}, and a practical example is provided in Section \ref{sec_evaluation_data_model}.
\\
\hline
\textbf{FR05} &
An \gls{EMS} must be able to fit system-specific parameters of a service utilizing its \gls{API}. &
An \gls{EMS} can fit system-specific parameters of a service (that utilizes the service framework) by interacting with the \texttt{/fit-parameters/} endpoints defined in Section \ref{sec_design_concept_api_design}.
\\
\hline
\textbf{FR06} &
A service developer must be able to specify the format of the input and output data exchanged while an \gls{EMS} interacts with the \gls{API} of a service to fit the system-specific parameters. &
A service developer can specify the format of the input and output data for the \texttt{/fit-parameters/} endpoints by defining the corresponding data model. The concept is defined in Section \ref{sec_design_concept_service_components}, and a practical example is provided in Section \ref{sec_evaluation_data_model}.
\\
\hline
\textbf{FR07} &
An \gls{EMS} must have the option to store fitted system-specific parameters locally. &
An \gls{EMS} can fetch the system-specific parameters (i.e. the output of the interaction with the \texttt{/fit-parameters/} endpoints) after the service (that utilizes the service framework) has finished the fitting process, see Section \ref{sec_design_concept_api_design}.
In fact, the service framework does not provide a functionality that would allow services to permanently store the fitted parameters. Further details are provided in Appendix \ref{appendix_interaction_patterns}. 
\\
\hline
\textbf{FR08} &
An \gls{EMS} must be able to make calls to the \gls{API} of the service which may take several hours to compute. &
An \gls{EMS} can make calls that take several hours (or much longer) to compute to the \texttt{/request/} and \texttt{/fit-parameters/} endpoints of a service (that utilizes the service framework) by first posting the demand for computation, then calling the respective \texttt{/status/} endpoint and finally fetching the result from the respective \texttt{/result/} endpoint, see Section \ref{sec_design_concept_api_design} for details.
\\
\hline
\textbf{FR09} &
A service developer should be able to automatically generate a documentation for the \gls{API} of a service. &
A service developer can automatically generate an interactive documentation for the \gls{API} of a service-based on the data model. See Section \ref{sec_implementation_api} for details.
\\
\hline
\textbf{NFR01} &
A service provider must be able to operate services with high availability and scalability. &
A service provider can operate services (that utilize the service framework) with high availability and scalability as the operation concept (see Section \ref{sec_design_concept_operation_concept}) for services explicitly considers execution on clusters.
Furthermore, the internal architecture of the services (see Section \ref{sec_design_concept_service_architecture}) supports scaling of \gls{API} and worker processes over multiple machines, an important preliminary for high availability and performance on clusters.
\\
\hline
\textbf{NFR02} &
An \gls{EMS} must be able to communicate with the service over an encrypted connection. &
An \gls{EMS} can communicate with services over an encrypted connection as the operation concept (see Section \ref{sec_design_concept_operation_concept}) demands that the service provider operates a gateway application in front of any service.
The gateway encrypts the communication between client and service with the \gls{HTTPS} protocol.
\\
\hline
\textbf{NFR03} &
A service provider must be able to restrict access to a service to authorized \glspl{EMS}. &
A service provider can restrict access to a service (that utilizes the service framework) by configuring the latter to verify that incoming calls contain a valid \gls{JWT}. A general discussion of this concept can be found in Section \ref{sec_design_concept_api_design}, while further implementation details and configuration options are provided in the online documentation of the service framework.
\\
\hline
\textbf{NFR04} &
A service developer/provider should be able to validate the correct implementation of the service framework. &
A service developer/provider can validate the correct implementation of the service framework as it is published as open-source repository, see Section \ref{sec_community_concept} for details.
\\
\hline
\textbf{NFR05} &
A service developer/provider should be able to verify that the service framework is actively maintained. &
A service developer/provider can verify that the service framework is actively maintained by checking the website of the Open Energy Services community or directly contacting members of the latter, see Section \ref{sec_community_concept} for details.
\\
\hline
\textbf{NFR06} &
A service developer should be able to derive a service with minimal effort from an existing forecasting or optimization algorithm. &
Utilizing the service framework, a service developer can derive a functional service with minimal effort as all functionality generic to services in general is concentrated in the implementation of the service framework, which is readily provided.
See Section \ref{sec_design_concept_service_components} for conceptual details as well as Section \ref{sec_evaluation_example} for an example demonstrating the implementation of a simple \gls{PV} power forecast service.
\\
\hline
\textbf{NFR07} &
A service provider should be able to operate a service without requiring expert knowledge about IT infrastructure. &
A service provider can operate a service without requiring expert knowledge about IT infrastructure as this is explicitly considered in the operation concept (see Section \ref{sec_design_concept_operation_concept}).
Concrete measures include support for Docker Swarm as orchestration system as well as the possibility to use an \gls{IdP} of an allied service provider.
\\
\hline
\textbf{NFR08} &
An \gls{EMS} developer must be able to specify which version of a service should be utilized. &
An \gls{EMS} developer can specify which version of a service should be utilized by filling in the desired version in the \texttt{\{version\}} placeholder that is a part of all \gls{API} endpoints, see Section \ref{sec_design_concept_api_design} for details.
\\
\hline
\textbf{NFR09} &
An \gls{EMS} developer should be able to quickly understand the \gls{API} of a service by utilizing the documentation. &
An \gls{EMS} developer can inspect the automatically generated interactive documentation, the gold standard regarding comprehensibility, of any service (that utilizes the service framework). 
See Section \ref{sec_implementation_api} for details.
\\
\hline
\textbf{NFR10} &
An \gls{EMS} developer should be able to implement a client to interact with the \gls{API} of a service with minimal effort. &
An \gls{EMS} developer can use the generic service client provided as part of the \gls{ESG} framework, see Section \ref{sec_evaluation_ease_of_client} for a practical example. As an alternative, it is possible to automatically generate the \gls{API}-related code of a client using the freely available Swagger Codegen tool for any service (that utilizes the service framework). 
See Section \ref{sec_implementation_api} for details.
\\
\hline
\end{longtable}
\section{Conclusion and Outlook}
The aim of this paper is to support the widespread adoption of \glspl{EMS} in order to unlock flexibility and energy savings potentials of end consumers.
We claim that economic viability is a severe issue for the utilization of \glspl{EMS} at scale and that the provisioning of forecasting and optimization algorithms as a service can make a major contribution to achieving it.
To this end, we introduce a software framework that allows the derivation of fully functional services from existing forecasting or optimization code with ease.
Our development of this framework is strictly systematic and begins by deducing requirements from an extensive analysis of the application of \glspl{EMS} in several domains.
Based on this, we derive a holistic design concept for the framework, covering the components, the architecture, and the operation of services.
We derive the \gls{ESG} package from the proposed design concept, and we publish it as free and open-source software alongside this work.
Beyond the service framework, this paper furthermore marks the starting point of the \emph{Open Energy Service} community, our effort to continuously maintain the service framework but also provide ready-to-use forecasting and optimization services, to bootstrap future research projects and to accelerate the widespread adoption of \glspl{EMS}.
This community is open for others to join, and any interest in participating is appreciated.
Finally, we demonstrate that our framework and our community concept are valuable contributions that meet the goals of this work.
To this end, we provide practical examples for the implementation of a service based on a simple \gls{PV} power generation forecasting code, as well as the corresponding client.
Furthermore, we demonstrate that our framework is capable of supporting the operation of services at relevant scales for \gls{EMS} applications.
We thus conclude that this work is a relevant step forward towards unlocking the potentials of forecasting and optimization algorithms provided as services for \glspl{EMS}, which hopefully supports the utilization of energy management applications at scale.
\begin{Backmatter}

\paragraph{Acknowledgments}
We would like to thank the anonymous reviewers for their constructive feedback, which helped us to improve this paper.
We furthermore like to thank Antonia Dieterich, who supported our analysis of related work.

\paragraph{Funding Statement}
This research has partly been funded by the German Federal Ministry for Economic Affairs and Climate Action within the projects FlexBlue and AMAZING, the Helmholtz Association under the Program Energy System Design  as well as the European Union within the project WeForming.\\
\includegraphics[width=0.30\linewidth]{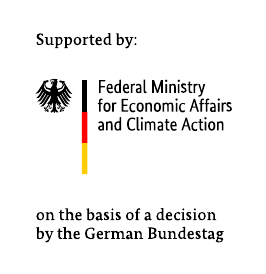}

\includegraphics[width=0.25\linewidth]{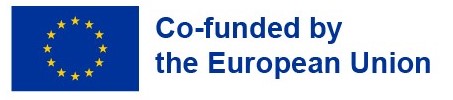}

\paragraph{Competing Interests}
None.

\paragraph{Data Availability Statement}
None.

\paragraph{Ethical Standards}
The research meets all ethical guidelines, including adherence to the legal requirements of the study country.

\paragraph{Author Contributions}
Conceptualization: D.W; K.F; R.M. Funding acquisition: D.W. T.R; V.H; H.S. Methodology: D.W; L.L. Project administration: D.W. Software: D.W. Supervision: R.M; V.H; H.S; Validation: D.W. Writing – original draft: D.W; K.F; T.R; N.F. Writing – review \& editing: D.W; K.F; T.R; N.F; R.M; V.H; H.S.
All authors approved the final submitted draft.

\paragraph{Supplementary Material}\label{sec_supplementary_material}
The \emph{Energy Service Generics} repository, including the source code as well as extensive documentation, can be found at:

\noindent \href{https://github.com/fzi-forschungszentrum-informatik/energy-service-generics}{https://github.com/fzi-forschungszentrum-informatik/energy-service-generics}

\noindent Further details and latest news about the \emph{Open Energy Services} community are provided on the corresponding website:

\noindent \href{https://open-energy-services.org/}{https://open-energy-services.org/}
\printnoidxglossary[type=acronym,sort=letter,nonumberlist,title={List of Abbreviations}]

\bibliographystyle{apalike}
\bibliography{library_dwo.bib,library_rm.bib,library_kfo.bib,library_nfe.bib}

\begin{thebibliography}{}

\bibitem[Al-Ghaili et~al., 2021]{al2021energy}
Al-Ghaili, A.~M., Kasim, H., Al-Hada, N.~M., Jørgensen, B.~N., Othman, M., and
  Wang, J. (2021).
\newblock {{Energy Management Systems and Strategies in Buildings Sector: A
  Scoping Review}}.
\newblock {\em IEEE Access}, 9:63790--63813.

\bibitem[Alebrahim et~al., 2014]{alebrahim2014structured}
Alebrahim, A., Heisel, M., and Meis, R. (2014).
\newblock {{A Structured Approach for Eliciting, Modeling, and Using
  Quality-Related Domain Knowledge}}.
\newblock In {\em Computational Science and Its Applications -- ICCSA 2014},
  pages 370--386, Cham. Springer International Publishing.

\bibitem[Alizadeh et~al., 2016]{alizadeh_flexibility_2016}
Alizadeh, M., Parsa~Moghaddam, M., Amjady, N., Siano, P., and
  {Sheikh-El-Eslami}, M. (2016).
\newblock Flexibility in future power systems with high renewable penetration:
  {{A}} review.
\newblock {\em Renewable and Sustainable Energy Reviews}, 57:1186--1193.

\bibitem[Alves et~al., 2006]{alves_study_2006}
Alves, C.~F., Pereira, S., and {de Castro}, J.~B. (2006).
\newblock A study in market-driven requirements engineering.
\newblock In {\em {{WER}}}, pages 61--66.

\bibitem[Anand et~al., 2023]{anand_bottom-up_2023}
Anand, H., Nateghi, R., and Alemazkoor, N. (2023).
\newblock Bottom-up forecasting: {{Applications}} and limitations in load
  forecasting using smart-meter data.
\newblock {\em Data-Centric Engineering}, 4:e14.

\bibitem[Anthi et~al., 2018]{anthi_secure_2018}
Anthi, E., Javed, A., Rana, O., and Theodorakopoulos, G. (2018).
\newblock {{Secure Data Sharing and Analysis in Cloud-Based Energy Management
  Systems}}.
\newblock In {\em Cloud Infrastructures, Services, and IoT Systems for Smart
  Cities}, pages 228--242, Cham. Springer International Publishing.

\bibitem[Antonelli et~al., 2012]{antonelli2012deriving}
Antonelli, L., Rossi, G., do~Prado~Leite, J. C.~S., and Oliveros, A. (2012).
\newblock Deriving requirements specifications from the application domain
  language captured by {{Language Extended Lexicon}}.
\newblock In {\em Proc. Workshop in Requirements Engineering (WER)}, pages
  24--27.

\bibitem[Berkes and Keshav, 2024]{berkes2024sopevs}
Berkes, A. and Keshav, S. (2024).
\newblock {{SOPEVS: Sizing and Operation of PV-EV-Integrated Modern Homes}}.
\newblock In {\em Proceedings of the 15th ACM International Conference on
  Future and Sustainable Energy Systems}, e-Energy '24, page 14–26, New York,
  NY, USA. Association for Computing Machinery.

\bibitem[Blechmann et~al., 2023]{blechmann2023open}
Blechmann, S., Sowa, I., Schraven, M.~H., Streblow, R., Müller, D., and Monti,
  A. (2023).
\newblock Open source platform application for smart building and smart grid
  controls.
\newblock {\em Automation in Construction}, 145:104622.

\bibitem[Boiko et~al., 2024]{boiko_edge-cloud_2024}
Boiko, O., Komin, A., Malekian, R., and Davidsson, P. (2024).
\newblock {{Edge-Cloud Architectures for Hybrid Energy Management Systems: A
  Comprehensive Review}}.
\newblock {\em IEEE Sensors Journal}, 24(10):15748--15772.

\bibitem[Chawda et~al., 2020]{chawda2020comprehensive}
Chawda, G.~S., Shaik, A.~G., Shaik, M., Padmanaban, S., Holm-Nielsen, J.~B.,
  Mahela, O.~P., and Kaliannan, P. (2020).
\newblock {Comprehensive Review on Detection and Classification of Power
  Quality Disturbances in Utility Grid With Renewable Energy Penetration}.
\newblock {\em IEEE Access}, 8:146807--146830.

\bibitem[Chen et~al., 2019]{chen_gnu-rl_2019}
Chen, B., Cai, Z., and Berg{\'e}s, M. (2019).
\newblock Gnu-{{RL}}: {{A Precocial Reinforcement Learning Solution}} for
  {{Building HVAC Control Using}} a {{Differentiable MPC Policy}}.
\newblock In {\em Proceedings of the 6th {{ACM International Conference}} on
  {{Systems}} for {{Energy-Efficient Buildings}}, {{Cities}}, and
  {{Transportation}}}, {{BuildSys}} '19, pages 316--325, New York, NY, USA.
  Association for Computing Machinery.

\bibitem[Cirillo et~al., 2019]{cirillo2019aStandard}
Cirillo, F., Solmaz, G., Berz, E.~L., Bauer, M., Cheng, B., and Kovacs, E.
  (2019).
\newblock {A Standard-Based Open Source IoT Platform: FIWARE}.
\newblock {\em IEEE Internet of Things Magazine}, 2(3):12--18.

\bibitem[Coelho and Valente, 2017]{coelho2017WhyModernOpen}
Coelho, J. and Valente, M.~T. (2017).
\newblock {{Why Modern Open Source Projects Fail}}.
\newblock In {\em Proceedings of the 2017 11th {{Joint Meeting}} on
  {{Foundations}} of {{Software Engineering}}}, {{ESEC}}/{{FSE}} 2017, pages
  186--196, {New York, NY, USA}. {Association for Computing Machinery}.

\bibitem[{Dawson-Haggerty} et~al., 2013]{dawson-haggerty_boss_2013}
{Dawson-Haggerty}, S., Krioukov, A., Taneja, J., Karandikar, S., Fierro, G.,
  Kitaev, N., and Culler, D. (2013).
\newblock {{BOSS}}: {{Building Operating System Services}}.
\newblock In {\em 10th {{USENIX}} Symposium on Networked Systems Design and
  Implementation ({{NSDI}} 13)}, pages 443--457, Lombard, IL. USENIX
  Association.

\bibitem[De~Jongh et~al., 2022]{de_jongh_physics-informed_2022}
De~Jongh, S., Gielnik, F., Mueller, F., Schmit, L., Suriyah, M., and Leibfried,
  T. (2022).
\newblock Physics-informed geometric deep learning for inference tasks in power
  systems.
\newblock {\em Electric Power Systems Research}, 211:108362.

\bibitem[Dengler et~al., 2023]{dengler_p4_2023}
Dengler, G., Lalbakhsh, P., Bazan, P., Dayaratne, T., Liebmann, A., and German,
  R. (2023).
\newblock P4 {{Poster}} abstract: A flexible simulation-optimization framework
  for smart grids using distributed agents.
\newblock In {\em Abstracts of the 12th {{DACH}}+ Conference on Energy
  Informatics 2023}. Springer.

\bibitem[Ding et~al., 2019]{ding_octopus_2019}
Ding, X., Du, W., and Cerpa, A. (2019).
\newblock {{OCTOPUS}}: {{Deep Reinforcement Learning}} for {{Holistic Smart
  Building Control}}.
\newblock In {\em Proceedings of the 6th {{ACM International Conference}} on
  {{Systems}} for {{Energy-Efficient Buildings}}, {{Cities}}, and
  {{Transportation}}}, {{BuildSys}} '19, pages 326--335, New York, NY, USA.
  Association for Computing Machinery.

\bibitem[Eckert et~al., 2019]{eckert2019AloneTogetherInterorganizational}
Eckert, R., Stuermer, M., and Myrach, T. (2019).
\newblock Alone or {{Together}}? {{Inter-organizational Affiliations of Open
  Source Communities}}.
\newblock {\em Journal of Systems and Software}, 149:250--262.

\bibitem[{Eurostat}, 2023]{eurostat_energy_2023}
{Eurostat} (2023).
\newblock Energy consumption in households.
\newblock
  https://ec.europa.eu/eurostat/statistics-explained/index.php?title=Energy\_consumption\_in\_households.

\bibitem[Fielding, 2000]{fielding_architectural_2000}
Fielding, R.~T. (2000).
\newblock {\em Architectural {{Styles}} and the {{Design}} of {{Network-based
  Software Architectures}}}.
\newblock PhD thesis, University of California, Irvine.

\bibitem[Fogel, 2022]{fogel_producing_2022}
Fogel, K. (2022).
\newblock Producing {{Open Source Software}}.
\newblock https://producingoss.com/en/index.html.

\bibitem[F\"{o}rderer et~al., 2022]{foerderer2022automated}
F\"{o}rderer, K., Hagenmeyer, V., and Schmeck, H. (2022).
\newblock {{Automated Generation of Models for Demand Side Flexibility Using
  Machine Learning: An Overview}}.
\newblock {\em SIGENERGY Energy Inform. Rev.}, 1(1):107–120.

\bibitem[Galenzowski et~al., 2023]{Galenzowski2023}
Galenzowski, J., Waczowicz, S., Meisenbacher, S., Mikut, R., and Hagenmeyer, V.
  (2023).
\newblock A real-world district community platform as a cyber-physical-social
  infrastructure systems in the energy domain.
\newblock In {\em Proceedings of the 10th ACM International Conference on
  Systems for Energy-Efficient Buildings, Cities, and Transportation}, pages
  434--441.

\bibitem[Glinz, 2007]{glinz_non-functional_2007}
Glinz, M. (2007).
\newblock On {{Non-Functional Requirements}}.
\newblock In {\em 15th {{IEEE International Requirements Engineering
  Conference}} ({{RE}} 2007)}, pages 21--26, Delhi. IEEE.

\bibitem[Gwerder et~al., 2013]{gwerder_final_2013}
Gwerder, M., Gyalistras, D., Sagerschnig, C., Smith, R.~S., and Sturzenegger,
  D. (2013).
\newblock Final {{Report}}: {{Use}} of {{Weather And Occupancy Forecasts For
  Optimal Building Climate Control Part II}}: {{Demonstration}}
  ({{OptiControl-II}}).
\newblock Technical report, ETH Z{\"u}rich.

\bibitem[Han et~al., 2023]{han_home_2023}
Han, B., Zahraoui, Y., Mubin, M., Mekhilef, S., Seyedmahmoudian, M., and
  Stojcevski, A. (2023).
\newblock Home {{Energy Management Systems}}: {{A Review}} of the {{Concept}},
  {{Architecture}}, and {{Scheduling Strategies}}.
\newblock {\em IEEE Access}, 11:19999--20025.

\bibitem[{Henggeler Antunes} et~al., 2022]{antunes2022comprehensive}
{Henggeler Antunes}, C., Alves, M.~J., and Soares, I. (2022).
\newblock A comprehensive and modular set of appliance operation {{MILP}}
  models for demand response optimization.
\newblock {\em Applied Energy}, 320:119142.

\bibitem[Hill et~al., 2023]{hill_p2_2023}
Hill, A., Pieper, C., Bruhn, J.-H., Sch{\"o}nfeldt, P., and Penaherrera~Vaca,
  F.~A. (2023).
\newblock P2 {{Poster}} abstract: District energy management simulation
  framework with rolling horizon approach.
\newblock In {\em Abstracts of the 12th {{DACH}}+ Conference on Energy
  Informatics 2023}. Springer.

\bibitem[Hofmeister et~al., 2024]{hofmeister_semantic_2024}
Hofmeister, M., Bai, J., Brownbridge, G., Mosbach, S., Lee, K.~F., Farazi, F.,
  Hillman, M., Agarwal, M., Ganguly, S., Akroyd, J., and Kraft, M. (2024).
\newblock Semantic agent framework for automated flood assessment using dynamic
  knowledge graphs.
\newblock {\em Data-Centric Engineering}, 5:e14.

\bibitem[Hunter, 2017]{hunter_irresistible_2017}
Hunter, K.~L. (2017).
\newblock {\em Irresistible {{APIs}}: Designing Web {{APIs}} That Developers
  Will Love}.
\newblock Manning, Shelter Island, NY.

\bibitem[IEEE, 2002]{ieee_glossary_2002}
IEEE (2002).
\newblock {{IEEE Standard Glossary}} of {{Software Engineering Terminology}}.

\bibitem[{IPCC}, 2022]{shukla_summary_2022}
{IPCC} (2022).
\newblock Summary for {{Policymakers}}.
\newblock In Shukla, P., Skea, J., Reisinger, A., Slade, R., Fradera, R.,
  Pathak, M., Khourdajie, A.~A., Belkacemi, M., {van Diemen}, R., Hasija, A.,
  Lisboa, G., Luz, S., Malley, J., McCollum, D., Some, S., and Vyas, P.,
  editors, {\em Climate {{Change}} 2022: {{Mitigation}} of {{Climate Change}}.
  {{Contribution}} of {{Working Group III}} to the {{Sixth Assessment Report}}
  of the {{Intergovernmental Panel}} on {{Climate Change}}}, pages 3--48.
  Cambridge University Press, 1 edition.

\bibitem[ISO/IEC, 2023]{iso25010}
ISO/IEC (2023).
\newblock {\em ISO/IEC 25010: Systems and software engineering —Systems and
  software Quality Requirements and Evaluation (SQuaRE) — Product quality
  model}.
\newblock ISO copyright office.

\bibitem[Jawarneh et~al., 2019]{jawarneh_container_2019}
Jawarneh, I. M.~A., Bellavista, P., Bosi, F., Foschini, L., Martuscelli, G.,
  Montanari, R., and Palopoli, A. (2019).
\newblock Container {{Orchestration Engines}}: {{A Thorough Functional}} and
  {{Performance Comparison}}.
\newblock In {\em {{ICC}} 2019 - 2019 {{IEEE International Conference}} on
  {{Communications}} ({{ICC}})}, pages 1--6.

\bibitem[Jin et~al., 2018]{jin_designing_2018}
Jin, B., Sahni, S., and Shevat, A. (2018).
\newblock {\em Designing {{Web APIs}}: Building {{APIs}} That Developers Love}.
\newblock O'Reilly, Beijing Boston Farnham, first edition.

\bibitem[Jones et~al., 2015]{jones_json_2015}
Jones, M.~B., Bradley, J., and Sakimura, N. (2015).
\newblock {{JSON}} web token ({{JWT}}).
\newblock {{RFC}} 7519, RFC Editor.

\bibitem[Khalid, 2024]{khalid2024smart}
Khalid, M. (2024).
\newblock Smart grids and renewable energy systems: Perspectives and grid
  integration challenges.
\newblock {\em Energy Strategy Reviews}, 51:101299.

\bibitem[Kondziella and Bruckner, 2016]{kondziella_flexibility_2016}
Kondziella, H. and Bruckner, T. (2016).
\newblock Flexibility requirements of renewable energy based electricity
  systems -- a review of research results and methodologies.
\newblock {\em Renewable and Sustainable Energy Reviews}, 53:10--22.

\bibitem[Kornienko et~al., 2021]{kornienko_principles_2021}
Kornienko, D.~V., Mishina, S.~V., Shcherbatykh, S.~V., and Melnikov, M.~O.
  (2021).
\newblock Principles of securing {{RESTful API}} web services developed with
  python frameworks.
\newblock {\em Journal of Physics: Conference Series}, 2094(3):032016.

\bibitem[Kotilainen, 2019]{Kotilainen2019}
Kotilainen, K. (2019).
\newblock {\em Energy Prosumers' Role in the Sustainable Energy System}, pages
  1--14.
\newblock Springer International Publishing, Cham.

\bibitem[Krebs and Cruz~Martinez, 2022]{krebs_developing_2022}
Krebs, B. and Cruz~Martinez, J. (2022).
\newblock Developing {{RESTful APIs}} with {{Python}} and {{Flask}}.

\bibitem[Kursawe et~al., 2011]{kursawe_privacy_2011}
Kursawe, K., Danezis, G., and Kohlweiss, M. (2011).
\newblock {{Privacy-Friendly Aggregation for the Smart-Grid}}.
\newblock In {\em Privacy Enhancing Technologies: 11th International Symposium,
  PETS 2011, Waterloo, ON, Canada, July 27-29, 2011. Proceedings 11}, pages
  175--191. Springer.

\bibitem[Lamprecht et~al., 2020]{lamprecht_towards_2020}
Lamprecht, A.-L., Garcia, L., Kuzak, M., Martinez, C., Arcila, R., Martin
  Del~Pico, E., Dominguez Del~Angel, V., Van De~Sandt, S., Ison, J., Martinez,
  P.~A., McQuilton, P., Valencia, A., Harrow, J., Psomopoulos, F., Gelpi,
  J.~L., Chue~Hong, N., Goble, C., and {Capella-Gutierrez}, S. (2020).
\newblock Towards {{FAIR}} principles for research software.
\newblock {\em Data Science}, 3(1):37--59.

\bibitem[Langer et~al., 2013]{langer2013privacy}
Langer, L., Skopik, F., Kienesberger, G., and Li, Q. (2013).
\newblock Privacy issues of smart e-mobility.
\newblock In {\em IECON 2013 - 39th Annual Conference of the IEEE Industrial
  Electronics Society}, pages 6682--6687.

\bibitem[Lee et~al., 2016]{lee_design_2016}
Lee, E.-K., Shi, W., Gadh, R., and Kim, W. (2016).
\newblock Design and {{Implementation}} of a {{Microgrid Energy Management
  System}}.
\newblock {\em Sustainability}, 8(11):1143.

\bibitem[Lenk et~al., 2020]{Lenk2020}
Lenk, S., Arnoldt, A., R{\"o}sch, D., and Bretschneider, P. (2020).
\newblock Hardware/software architecture to investigate resilience in energy
  management for smart grids.
\newblock In {\em 2020 IEEE PES Innovative Smart Grid Technologies Europe
  (ISGT-Europe)}, pages 51--55. IEEE.

\bibitem[Loucopoulos and Champion, 1988]{loucopoulos1988knowledge}
Loucopoulos, P. and Champion, R. (1988).
\newblock Knowledge-based approach to requirements engineering using method and
  domain knowledge.
\newblock {\em Knowledge-Based Systems}, 1(3):179--187.

\bibitem[Malviya and Dwivedi, 2022]{malviya_comparative_2022}
Malviya, A. and Dwivedi, R.~K. (2022).
\newblock A {{Comparative Analysis}} of {{Container Orchestration Tools}} in
  {{Cloud Computing}}.
\newblock In {\em 2022 9th {{International Conference}} on {{Computing}} for
  {{Sustainable Global Development}} ({{INDIACom}})}, pages 698--703.

\bibitem[Maree and Bagle, 2022]{maree_building_2022}
Maree, J.~P. and Bagle, M. (2022).
\newblock A {{Building Automation}} and {{Control}} micro-service architecture
  using {{Physics Inspired Neural Networks}}.
\newblock {\em E3S Web of Conferences}, 362:13001.

\bibitem[Marinakis et~al., 2020]{marinakis_big_2020}
Marinakis, V., Doukas, H., Tsapelas, J., Mouzakitis, S., Sicilia, {\'A}.,
  Madrazo, L., and Sgouridis, S. (2020).
\newblock From big data to smart energy services: {{An}} application for
  intelligent energy management.
\newblock {\em Future Generation Computer Systems}, 110:572--586.

\bibitem[{Mateos-Garcia} and Steinmueller,
  2008]{mateos-garcia_institutions_2008}
{Mateos-Garcia}, J. and Steinmueller, W.~E. (2008).
\newblock The institutions of open source software: {{Examining}} the
  {{Debian}} community.
\newblock {\em Information Economics and Policy}, 20(4):333--344.

\bibitem[Mauser et~al., 2015]{mauser_organic_2015}
Mauser, I., Hirsch, C., Kochanneck, S., and Schmeck, H. (2015).
\newblock Organic {{Architecture}} for {{Energy Management}} and {{Smart
  Grids}}.
\newblock In {\em 2015 {{IEEE International Conference}} on {{Autonomic
  Computing}}}, pages 101--108, Grenoble, France. IEEE.

\bibitem[Meisenbacher et~al., 2023]{meisenbacher_autopv_2023}
Meisenbacher, S., Heidrich, B., Martin, T., Mikut, R., and Hagenmeyer, V.
  (2023).
\newblock {{AutoPV}}: {{Automated}} photovoltaic forecasts with limited
  information using an ensemble of pre-trained models.
\newblock In {\em Proceedings of the 14th {{ACM International Conference}} on
  {{Future Energy Systems}}}, pages 386--414, Orlando FL USA. ACM.

\bibitem[Mercl and Pavlik, 2019]{yang_comparison_2019}
Mercl, L. and Pavlik, J. (2019).
\newblock The {{Comparison}} of {{Container Orchestrators}}.
\newblock In Yang, X.-S., Sherratt, S., Dey, N., and Joshi, A., editors, {\em
  Third {{International Congress}} on {{Information}} and {{Communication
  Technology}}}, volume 797, pages 677--685. Springer Singapore, Singapore.

\bibitem[Mohamed et~al., 2018]{mohamed_service-oriented_2018}
Mohamed, N., {Al-Jaroodi}, J., and Jawhar, I. (2018).
\newblock Service-{{Oriented Big Data Analytics}} for {{Improving Buildings
  Energy Management}} in {{Smart Cities}}.
\newblock In {\em 2018 14th {{International Wireless Communications}} \&
  {{Mobile Computing Conference}} ({{IWCMC}})}, pages 1243--1248, Limassol.
  IEEE.

\bibitem[Oldewurtel et~al., 2012]{oldewurtel_use_2012}
Oldewurtel, F., Parisio, A., Jones, C.~N., Gyalistras, D., Gwerder, M., Stauch,
  V., Lehmann, B., and Morari, M. (2012).
\newblock Use of model predictive control and weather forecasts for energy
  efficient building climate control.
\newblock {\em Energy and Buildings}, 45:15--27.

\bibitem[Papaefthymiou and Dragoon, 2016]{papaefthymiou_towards_2016}
Papaefthymiou, G. and Dragoon, K. (2016).
\newblock Towards 100\% renewable energy systems: {{Uncapping}} power system
  flexibility.
\newblock {\em Energy Policy}, 92:69--82.

\bibitem[Pipattanasomporn et~al., 2015]{pipattanasomporn_bemoss_2015}
Pipattanasomporn, M., Kuzlu, M., Khamphanchai, W., Saha, A., Rathinavel, K.,
  and Rahman, S. (2015).
\newblock {{BEMOSS}}: {{An Agent Platform}} to {{Facilitate Grid-Interactive
  Building Operation}} with {{IoT Devices}}.
\newblock In {\em 2015 {{IEEE Innovative Smart Grid Technologies}} - {{Asia}}
  ({{ISGT ASIA}})}, pages 1--6, Bangkok, Thailand. IEEE.

\bibitem[Pohl, 1996]{pohl_requirements_1996}
Pohl, K. (1996).
\newblock Requirements engineering: {{An}} overview.
\newblock In {\em Encyclopedia of {{Computer Science}} and {{Technology}}},
  volume 36 - supp. 21. CRC Press.

\bibitem[Regnell and Brinkkemper, 2005]{aurum_market-driven_2005}
Regnell, B. and Brinkkemper, S. (2005).
\newblock Market-{{Driven Requirements Engineering}} for {{Software Products}}.
\newblock In Aurum, A. and Wohlin, C., editors, {\em Engineering and {{Managing
  Software Requirements}}}, pages 287--308. Springer-Verlag, Berlin/Heidelberg.

\bibitem[Roccotelli et~al., 2022]{roccotelli2022smart}
Roccotelli, M., Mangini, A.~M., and Fanti, M.~P. (2022).
\newblock {{Smart District Energy Management With Cooperative Microgrids}}.
\newblock {\em IEEE Access}, 10:36311--36326.

\bibitem[Rodriguez et~al., 2018]{rodriguez2018fiware}
Rodriguez, M.~A., Cuenca, L., and Ortiz, A. (2018).
\newblock {FIWARE Open Source Standard Platform in Smart Farming - A Review}.
\newblock In Camarinha-Matos, L.~M., Afsarmanesh, H., and Rezgui, Y., editors,
  {\em Collaborative Networks of Cognitive Systems}, pages 581--589, Cham.
  Springer International Publishing.

\bibitem[Ruhnau et~al., 2019]{ruhnau2019direct}
Ruhnau, O., Bannik, S., Otten, S., Praktiknjo, A., and Robinius, M. (2019).
\newblock {{Direct or indirect electrification? A review of heat generation and
  road transport decarbonisation scenarios for Germany 2050}}.
\newblock {\em Energy}, 166:989--999.

\bibitem[Saeed and Abdallah, 2022]{saeed_security_2022}
Saeed, L. and Abdallah, G. (2022).
\newblock {\em Security with {{JWT}}}, pages 293--308.
\newblock Apress, Berkeley, CA.

\bibitem[Salpakari and Lund, 2016]{salpakari_optimal_2016}
Salpakari, J. and Lund, P. (2016).
\newblock Optimal and rule-based control strategies for energy flexibility in
  buildings with {{PV}}.
\newblock {\em Applied Energy}, 161:425--436.

\bibitem[Schibuola et~al., 2015]{schibuola_demand_2015}
Schibuola, L., Scarpa, M., and Tambani, C. (2015).
\newblock Demand response management by means of heat pumps controlled via real
  time pricing.
\newblock {\em Energy and Buildings}, 90:15--28.

\bibitem[Scott and Neray, 2021]{scott_best_2021}
Scott, S. and Neray, G. (2021).
\newblock Best practices for {{REST API}} security: {{Authentication}} and
  authorization.
\newblock
  https://stackoverflow.blog/2021/10/06/best-practices-for-authentication-and-authorization-for-rest-apis/.

\bibitem[Shaikh et~al., 2014]{shaikh_review_2014}
Shaikh, P.~H., Nor, N. B.~M., Nallagownden, P., Elamvazuthi, I., and Ibrahim,
  T. (2014).
\newblock A review on optimized control systems for building energy and comfort
  management of smart sustainable buildings.
\newblock {\em Renewable and Sustainable Energy Reviews}, 34:409--429.

\bibitem[Singh, 2023]{singh_flask_2023}
Singh, R. (2023).
\newblock Flask vs {{FastAPI}}: {{Which Python Web Framework}} is {{Right}} for
  {{You}}?
\newblock
  https://www.linkedin.com/pulse/flask-vs-fastapi-which-python-web-framework-right-you-ritwik-singh-ddxtc/.

\bibitem[Sovacool et~al., 2022]{SOVACOOL2022112868}
Sovacool, B.~K., Barnacle, M.~L., Smith, A., and Brisbois, M.~C. (2022).
\newblock Towards improved solar energy justice: Exploring the complex
  inequities of household adoption of photovoltaic panels.
\newblock {\em Energy Policy}, 164:112868.

\bibitem[Sp{\"a}th, 2023]{spath_rest_2023}
Sp{\"a}th, P. (2023).
\newblock {\em {{REST Security}}}, pages 175--194.
\newblock Apress, Berkeley, CA.

\bibitem[Srilakshmi and Singh, 2022]{srilakshmi2022energy}
Srilakshmi, E. and Singh, S.~P. (2022).
\newblock {{Energy regulation of EV using MILP for optimal operation of
  incentive based prosumer microgrid with uncertainty modelling}}.
\newblock {\em International Journal of Electrical Power \& Energy Systems},
  134:107353.

\bibitem[Varenhorst et~al., 2024]{varenhorst2024enhancing}
Varenhorst, I. A.~M., Hoogsteen, G., Gerards, M. E.~T., and Hurink, J.~L.
  (2024).
\newblock {{Enhancing Privacy Through Time Aggregation of Load Profiles in
  Energy Management}}.
\newblock In {\em 2024 IEEE 8th Energy Conference (ENERGYCON)}, pages 1--6.

\bibitem[Volk et~al., 2017]{volk2017grid}
Volk, K., Lakenbrink, C., Kurka, C., and Rupp, L. (2017).
\newblock {{Grid-Control - An Overall Concept for the Distribution Grid of the
  "Energiewende"}}.
\newblock In {\em International ETG Congress 2017}, pages 1--6.

\bibitem[Washizaki, 2024]{washizaki_guide_2024}
Washizaki, H., editor (2024).
\newblock {\em Guide to the Software Engineering Body of Knowledge
  ({{SWEBOK}}): {{Version}} 4.0}.
\newblock IEEE Computer Society Press, Washington, DC, USA, 4rd edition.

\bibitem[W{\"o}lfle et~al., 2020]{wolfle_guide_2020}
W{\"o}lfle, D., Vishwanath, A., and Schmeck, H. (2020).
\newblock A {{Guide}} for the {{Design}} of {{Benchmark Environments}} for
  {{Building Energy Optimization}}.
\newblock In {\em Proceedings of the 7th {{ACM International Conference}} on
  {{Systems}} for {{Energy-Efficient Buildings}}, {{Cities}}, and
  {{Transportation}}}, {{BuildSys}} '20, pages 220--229, New York, NY, USA.
  Association for Computing Machinery.

\bibitem[Xuereb~Conti et~al., 2023]{xuereb_conti_physics-based_2023}
Xuereb~Conti, Z., Choudhary, R., and Magri, L. (2023).
\newblock A physics-based domain adaptation framework for modeling and
  forecasting building energy systems.
\newblock {\em Data-Centric Engineering}, 4:e10.

\bibitem[Zafar et~al., 2020]{zafar_hems_2020}
Zafar, U., Bayhan, S., and Sanfilippo, A. (2020).
\newblock {{Home Energy Management System Concepts, Configurations, and
  Technologies for the Smart Grid}}.
\newblock {\em IEEE Access}, 8:119271--119286.

\end{thebibliography}

\end{Backmatter}

\appendix

%
%
%
\newpage
\section{Interaction Patterns Between \gls{EMS} and Service}\label{appendix_interaction_patterns}
In Section \ref{sec_requirements_functional} we have discussed that it may be necessary to fit system-specific parameters to utilize the full potential of a forecasting or optimization service.
Furthermore, we have pointed out that this fitting process will usually need some form of historic measurements and that some \gls{EMS} users will want to keep these historic measurements on-premise while others will not be capable of doing so.
In this section, we will contrast the resulting interaction patterns between \gls{EMS} and service based on this important difference. To this end, we will return to the example used in the functional requirements section, i.e. a service that provides forecasts of \gls{PV} power generation.
\subsection{Local Storage of Measurements and Parameters}\label{appendix_interaction_patterns_local}
Let us first inspect the case of an \gls{EMS} that keeps historic measurements on-premise, i.e. a system with an architecture like the one displayed in Figure \ref{fig_ems_architecture_with_services}. The interaction of such an \gls{EMS} with the considered \gls{PV} power generation forecast service would contain the following steps:
\begin{enumerate}
	\item The \gls{EMS} calls the \gls{API} endpoint of the service related to the fitting process. This includes pushing historic measurements of the power generated by the \gls{PV} system for which forecasts should be computed. Note that the \gls{EMS} may need to repeat this step in a certain interval, especially once the forecasting performance deteriorates. Finding the right interval for refitting the parameters is solely the responsibility of the \gls{EMS}, as the service provider has no access to the measured data and can thus not evaluate the performance of the forecasts provided by the service.
	\item The \gls{EMS} retrieves the system-specific parameters once the service has finished the fitting process. The \gls{EMS} must store these parameters locally.
	\item The \gls{EMS} requests a forecast from the service by issuing a call to the respective \gls{API} endpoint of the service (this is likely repeated with arbitrary periodicity). The call must contain the arguments (e.g. the coordinates of the target system) as well as the fitted parameters (as returned in the previous step).
	\item The \gls{EMS} retrieves the computed forecast once the service has finished computing it.
\end{enumerate}
With this interaction pattern, the service cannot request any data of the \gls{HAL}, as the \gls{EMS} needs to actively push the data to the service. This means that the \gls{EMS} (developer) is always in control over the data stored by the system.
\subsection{Cloud Storage of Measurements and Parameters}\label{appendix_interaction_patterns_cloud}
\begin{figure}
\centering
\FIG{\includegraphics[width=1.0\columnwidth]{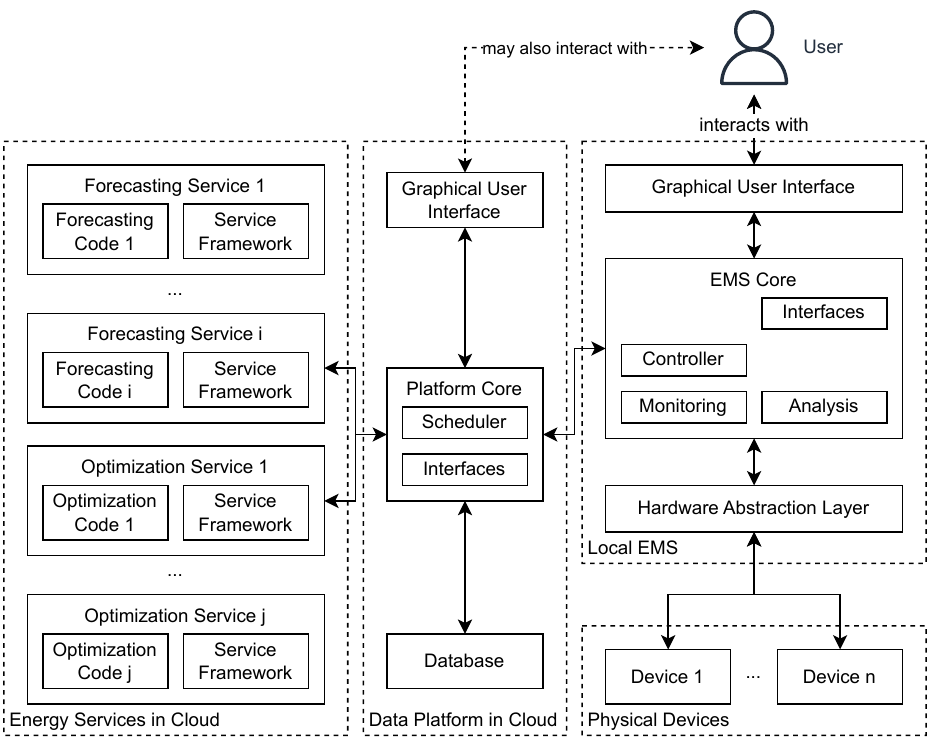}}{\caption{High-level architecture of an \gls{EMS} indirectly utilizing forecasting and optimization services via a platform. Note that some functionality like Monitoring or Analysis may, in fact, be implemented as part of the cloud data platform instead}\label{fig_ems_architecture_with_services_and_platform}}
\end{figure}
The second case is an \gls{EMS} that has been configured to push the relevant historic measurements to a data platform that is accessible to the service provider, as summarized in Figure \ref{fig_ems_architecture_with_services_and_platform}

This scenario is closer to a conventional "as a service" construct and allows the service provider to take over responsibility and complexity from the \gls{EMS}.
The price for this convenience is that the historic measurements of the \gls{EMS} are shared with the service provider, which might occur as a privacy issue for some users.
The resulting interaction pattern between the \gls{EMS}, data platform, and service contains the following steps:
\begin{enumerate}
	\item The \gls{EMS} continuously pushes all relevant measurements to the data platform.
	Furthermore, the \gls{EMS} must upload all other information required to invoke the desired services (e.g. the coordinates representing the global position of the target \gls{PV} system) to the data platform.
	\item The service provider\footnote{In fact this could be a third party that is not a service provider but invokes the services on behalf of the \gls{EMS}, e.g. as a commercial offering. However, this does not change the interaction pattern and is thus neglected here.} retrieves the relevant historic measurements from the data platform and uses these to call the \gls{API} endpoint of the service related to the fitting process. 
	\item The service provider retrieves the system-specific parameters after the service has finished computing these and stores the parameters in the data platform\footnote{Actually, it is not important whether the parameters are stored in the same data platform or some other database.}.
	\item At periodic intervals (e.g. every 15 minutes) or certain events (new information is available), the service provider retrieves the system-specific parameters as well as all other information to invoke the intended service and utilizes this data to request the forecast required by the \gls{EMS} from the service.
	\item Once the service has finished computing the forecast, the service provider collects it from the service and writes the forecast to the data platform.
	\item The \gls{EMS} periodically polls the data platform for new forecasts and retrieves these once available.
\end{enumerate}
\end{document}